%% file: main.tex
\documentclass[sigconf]{acmart}

\AtBeginDocument{%
  \providecommand\BibTeX{{%
    \normalfont B\kern-0.5em{\scshape i\kern-0.25em b}\kern-0.8em\TeX}}}

\copyrightyear{2023}
\acmYear{2023}
\setcopyright{rightsretained}
\acmConference[CHI '23]{Proceedings of the 2023 CHI Conference on Human Factors in Computing Systems}{April 23--28, 2023}{Hamburg, Germany}
\acmBooktitle{Proceedings of the 2023 CHI Conference on Human Factors in Computing Systems (CHI '23), April 23--28, 2023, Hamburg, Germany}
\acmDOI{10.1145/3544548.3581244}
\acmISBN{978-1-4503-9421-5/23/04}

\usepackage{subcaption}
\usepackage{multirow}

\newcommand{\sys}{\textit{NewsComp}}
\newcommand{\sysfull}{Comparative News Annotation}
\newcommand{\papertitle}{NewsComp: Facilitating Diverse News Reading through Comparative Annotation}

\newcommand{\rqt}[2]{\hspace{5pt} \textbf{RQ#1}. \textit{#2}}

\newcommand{\addition}[1]{#1}
\newcommand{\additionminor}[1]{#1}

\newcommand{\inlineqt}[2]{\textit{``#2''} - \textit{#1}}

\newcommand{\participants}{109}
\newcommand{\topics}{two}

\newcommand{\experts}{two}

\newcommand{\rqone}{How well do users perform comparative annotation?}

\newcommand{\rqtwo}{How does comparative news annotation affect users' perceptions of credibility and news quality?}

\begin{document}



\title[\sysfull]{\papertitle}


\settopmatter{authorsperrow=4}

\author{Md Momen Bhuiyan}
\authornote{The author conducted a portion of the work while interning at the University of Washington.}
\affiliation{%
  \institution{Virginia Tech}
  \city{Blacksburg}
  \country{USA}}
\email{momen@vt.edu}

\author{Sang Won Lee}
\affiliation{%
  \institution{Virginia Tech}
  \city{Blacksburg}
  \country{USA}}
\email{sangwonlee@vt.edu}

\author{Nitesh Goyal}
\affiliation{%
  \institution{Google Research}
  \city{New York}
  \country{USA}}
\email{niteshgoyal@acm.org}

\author{Tanushree Mitra}
\affiliation{%
  \institution{University of Washington}
  \city{Seattle}
  \country{USA}}
\email{tmitra@uw.edu}


\renewcommand{\shortauthors}{Md Momen Bhuiyan et al.}



\begin{abstract}
\addition{To support efficient, balanced news consumption, merging articles from diverse sources into one, potentially through crowdsourcing, could alleviate some hurdles.
However, the merging process could also impact annotators' attitudes towards the content.}
To test this theory, we propose comparative news annotation; that is, annotating similarities and differences between a pair of articles.
By developing and deploying \sys{}---a prototype system---we conducted a between-subjects experiment (N~=~\participants) to examine how users' annotations compare to experts', and how comparative annotation affects users' perceptions of article credibility and quality.
We found that comparative annotation can marginally impact users' credibility perceptions in certain cases; it did not impact perceptions of quality.
While users' annotations were not on par with experts', they showed greater precision in finding similarities than in identifying disparate important statements.
\addition{The comparison process also led users to notice differences in information placement and depth, degree of factuality/opinion, and empathetic/inflammatory language use.}
\addition{We discuss implications for the design of future comparative annotation tasks.}
\end{abstract}







\maketitle

\input{tex/2intro}
\input{tex/3related_work}

\input{tex/4method}
\input{tex/5result}
\input{tex/6discussion}

\input{tex/7conclusion}


\balance
\bibliographystyle{ACM-Reference-Format}
\bibliography{main}


\input{tex/8appendix}

\end{document}

%% file: tex/2intro.tex
\section{Introduction}
News media often produces content that is significantly biased in favor of a particular ideology, especially on contentious topics~\cite{herman2010manufacturing,mullainathan2002media,sunstein1999law}, and news consumers are affected by such biases~\cite{demarzo2003persuasion}.
Therefore, developing an informed opinion on a subject requires critically consuming news content from multiple sources.
While the internet gives users access to news from multiple sources,
when given choices, people tend to choose content that aligns with their viewpoints due to confirmation-seeking tendencies~\cite{sunstein2009http,nickerson1998confirmation,chhabra2013does,stroud2010polarization,nickerson1998confirmation}. 
Furthermore, the task of engaging with multiple perspectives is not easy and probably not performed equitably by all users~\cite{Howpeopl90online,Overview24online}.
\addition{One potential solution to this problem could be to use experts (i.e., journalists) to combine news items on an event from varying sources into a single story.
However, a limited number of experts would likely find it difficult to manage the volume of news stories generated by news outlets from around the world.
On the other hand, studies have shown that crowdworkers' output can be significantly correlated with experts' in some annotation tasks~\cite{allen2021scaling,allen2022birds,bhuiyan2020investigating,surowiecki2005wisdom,budescu2015identifying}.
Building on such results, this work explores whether crowdsourcing could be a viable approach to combining news articles from varying sources.
For a lay user, such a crowdsourcing task can be broken down into two aspects of comparative annotation: (i) finding similarities and (ii) finding important disparities.
These annotations can be useful to both news consumers and fact-checkers, whether professional or crowdsourced (e.g., BirdWatch\footnote{https://blog.twitter.com/en\_us/topics/product/2021/introducing-birdwatch-a-community-based-approach-to-misinformation}).
For everyday news consumers, merged articles can provide balanced perspectives on news events.
Second, fact-checkers can use similarity/dissimilarity annotations to validate claims through multiple sources or trace the origin of specific statements.
Besides, a by-product of any annotation task is that performing the task could also affect the annotators attitude towards the content, in our case, the news articles or the issue at hand.
In this work, we ask:}

\rqt{1}{\rqone{}}

\rqt{2}{\rqtwo{}}

Here, we use a simplified notion of comparative annotation: statements that are similar and statements that are dissimilar but important.
Using the concept of comparative news annotation, we developed and tested \sys{} (see the interface in Figure~\ref{fig:newscomp-interface}): a prototype that allows readers to compare and annotate similar and contrasting statements between only two news articles.
\sys{} has two components: (i) a comparative or side-by-side view of two articles from different sources, and (ii) an annotation tool.
Specifically, the annotation tool allows performing the two annotation tasks: (i) identifying similar statements across a pair of articles and connecting them with lines ({\includegraphics[scale=0.3]{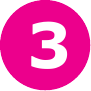}} 
 in Figure \ref{fig:newscomp-interface}), and (ii) identifying disparate statements (statements with no similarity) from each article that are important and should be included in the other article in the pair ({\includegraphics[scale=0.3]{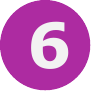}} 
 in Figure \ref{fig:newscomp-interface}).
\addition{To design the tool, we conducted a series of think-aloud formative studies with Google Drawings 
to observe the annotation process.
During those interviews, we noticed users considering different criteria for annotation.
For example, some considered only the content in each statement, while others considered the underlying themes behind the statements.}
Informed by the think-aloud sessions, we ask annotators to provide the reasoning behind their annotation (e.g., {\includegraphics[scale=0.3]{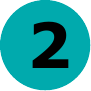}} 
  in Figure \ref{fig:newscomp-interface}).

To answer our research questions, we conducted a between-subjects experiment with \sys{} in a controlled environment.
We recruited \participants{} participants using Facebook advertising, which allowed us to recruit users from a large and diverse pool.
Participants were randomly assigned to either the treatment or the control group. 
For the study, we used two pairs of articles on \topics{} contentious topics: immigration and abortion.
To generate gold standards for the sake of comparison, we recruited \experts{} experts from the university's Department of Communication (one of whom had five years experience as a journalist) and asked them to perform annotation and rate the articles. 
Our experts found different degrees of similarity between the two pairs of articles; specifically, the pair of immigration articles had high dissimilarity (high contrast), while the pair of abortion articles pair was highly similar (low contrast).
During the study, users in the treatment read and annotated a pair of articles on the same topic, and then responded to a questionnaire designed to address our research questions (related to perceptions of credibility and quality). Meanwhile, users in the control group read a pair of articles on separate events without adding annotations and responded to the same questionnaire. 
We analyzed the extent to which news consumers' annotations matched experts' annotations, the impact of article topic and users' news expertise (knowledge of current events, perceived value of media literacy) on annotation quality, the reasons behind annotations, and how the treatment group's article perception compared to the control.
  
Regarding RQ1, we found that users performed poorly on both annotation tasks.
However, they had higher precision in finding similarities than in identifying important statements among the disparate statements.
We also found that filtering out annotations based on the number of users who annotated an item can rule out some false positives in finding similar statements, thus improving their collective F1 score.
In our study, ruling out annotations made by fewer than five or six users produced the highest F1 score.
Users with low current event knowledge made more annotations and had higher recall.
We also found that while annotating statement similarity, users provided different types of criteria, such as seeing connections when two statements discuss the same person, location, date, quote, or other information.
Among statements with no similarity, when annotating if a statement is important and should be included in the other article, users sometimes marked a statement important if it provided clarification or elaboration on other statements or if it provided a missing perspective.
Furthermore, we found that both generic words (e.g., ``quote'' and ``similar'') and article-specific words (e.g., ``lawsuit'') mentioned in the rationales can differentiate incorrect annotations from correct ones.
Perhaps such generic words in rationales can be used to filter out false positives in annotations on articles on different topics.
\addition{Comparing the articles, annotators also saw differences in perspectives presented, information placement, depth of detail, amount of factual/opinion statements, empathetic presentation, and use of inflammatory language.
Perceptions of \sys{} itself were mixed, though skewed more towards positive than negative.}
Regarding RQ2, we found that the treatment group's credibility ratings were significantly different compared to the control group's for high-contrast articles. For low-contrast articles, users in both groups performed similarly.
There were no significant effects on perceptions of quality.
Overall, this study indicates that we can leverage the comparative annotation mechanism to engage users in reading multiple perspectives.
However, since users produce annotations with high error rates, creating tools to assist in annotation could help reduce errors.
\addition{
We discuss applications for annotated data, such as developing a holistic view of an event from multiple news sources, teaching machines to discern article quality, training machine learning algorithms to generate better annotations, and assisting fact-checkers in their work.
We conclude with implications for the design of future comparative annotation tasks, such as modularizing into subtasks, providing supporting features to reduce load, and supporting co-annotation by multiple users.
}

%% file: tex/3related_work.tex
\section{Background and Related Works}
In this section, we begin by providing some background on media bias and multi-perspective online news consumption. Thereafter, we discuss related research on designing annotation tools for making sense of information and the effects of such annotations. 
 
\subsection{The Need for Multiperspective News Consumption}
While news articles should ideally follow established journalistic practices, various forms of biases and inaccuracies are injected into articles during the content production process.
This begins in the information gathering stage, where journalists must select events and related facts from sources.
In doing so, news publishers can influence which topics readers perceive to be relevant by selectively reporting on topics of their choosing~\cite{scheufele2000agenda}.
Next, journalists include and exclude information from sources (e.g., press releases, other news articles, and studies), shaping the perspective on the event.
In the writing phase, journalists make stylistic choices which may reflect their view of the news item, thereby producing biased coverage. 
For instance, journalists may introduce bias through the use of labeling (``a senator'' vs. ``a Republican senator'') and word choice (``illegal alien'' vs. ``undocumented immigrant'').
Such methods allow journalists to promote a particular interpretation of a topic~\cite{entman2007framing}.

Research suggests that a majority of news consumers are affected by media bias~\cite{napolitan1972election,demarzo2003persuasion,kull2003misperceptions} in different ways~\cite{druckman2005impact,scheufele2000agenda}.
Such bias can influence voting or election outcomes~\cite{napolitan1972election,druckman2005impact,dellavigna2007fox}.
Furthermore, media bias promotes polarization in public opinion, especially on contentious topics~\cite{sunstein1999law}. 
Some scholars argue that media bias challenges the pillars of American democracy~\cite{zaller1992nature,kahneman2013choices}.
Overall, these works point to the need to consume news from diverse perspectives to deal with biases in the media.

\subsection{Barriers to Multiperspective News Consumption Online}
Lazarsfeld et al. introduced the two-step flow model of communication, referring to the two gatekeeping stages that occur before an individual forms an opinion on a subject: first by news organizations, and then by opinion leaders in the individual's social circle~\cite{lazarsfeld1968people}.
Even though the internet has democratized access to information, including news, news consumption in the internet age still seems to follow the two-step flow model in communication in at least two ways: news selection and consumption~\cite{lazarsfeld1968people,choi2015two,soffer2021algorithmic}.
First, personalization algorithms act as filters for content selection; thus, they perform a gatekeeping function similar to that of opinion leaders in the pre-internet age~\cite{soffer2021algorithmic,pariser2011filter,stroud2011niche}.
Second, pervasive, echo chamber--esque news comment sections tend to promote opinions from opinion leaders with views aligned with users' own beliefs~\cite{jamieson2008echo,choi2015two}.
One problematic aspect of this internet-based, two-step communication is that users may not be aware of the second gatekeeping stage, given that algorithmic effects are often hidden, and partisan biases of opinion leaders in comment sections may also be obscured by anonymity~\cite{soffer2021algorithmic,correa2015many}.
Even when readers become aware of content with a political slant opposed to their own, they may lack the motivation to consume that content that due to political polarization and confirmation-seeking tendencies~\cite{center2014political,baker2021immigration,fiorina2008political,sunstein2009http,klayman1995varieties,nickerson1998confirmation,del2017modeling}.
Indeed, some research has found that while people might read more content when using diverse content selection tools, this leads primarily to an increase in the amount of content consumed, not the diversity of the content~\cite{chhabra2013does}.
One reason for this outcome could be individual differences between diversity-seeking and challenge-averse people; challenge-averse people may tend not to consume diverse content~\cite{munson2010presenting}.
To address this bias, some prior works developed mechanisms to promote diverse news selection through design tools, such as NewsCube~\cite{park2009newscube,Park2011,Park2011a,Park2012}, or through nudges to read alternative viewpoints~\cite{munson2013encouraging}.
Though these prior works demonstrated improved exposure---that is, clicks or visits to news sites with diverse political slants---there is a gap in our understanding of whether design tools can encourage critical engagement and whether such engagement affects users' perceptions of the news.
Furthermore, these tools do not ensure that people read articles on the same events from politically diverse sources.
We aim to bridge this gap by bundling pairs of articles on the same event from differing perspectives in a comparative annotation interface to test engagement and its effects.

\vspace*{-8pt}
\subsection{Designing for Information Consumption through Comparison}
Scholarship on reasoning, comprehension, and learning outlines different mechanisms in understanding information, whether users learn from data or the structure of information~\cite{arthur1994inductive,kavale1980reasoning,vosniadou1989similarity}.
Sometimes, reading multiple sources alone can help change a reader's mental model of a subject~\cite{stahl1996happens,braaten2011measuring}.
Comparison can further help people recognize common features shared across items or identify features that distinguish them~\cite{kavale1980reasoning,vosniadou1989similarity,vosniadou1987theories,chi1992conceptual}.
Some suggest that a comparison mechanism allows users to create broad concept categories by grouping similar concepts in either a bottom-up approach (clustering) or a top-down approach (assigning existing categories)~\cite{Zhang2020}.
In HCI research, creating design elements or affordances for easier information consumption is not new. For example, interactive elements allow users to choose where to go or what to read next~\cite{eveland2001user}. Such design elements can assist readers in constructing a cognitive model to support a thorough understanding of a news event.
This construction of a news schema is supported by providing signals---layouts, visual elements, and textual structures---to news readers that meet their expectations for news~\cite{Kiesow2021}.
For example, a newspaper reader's understanding of certain affordances (e.g., section labels, such as ``opinion'') may assist them in contextualizing and understanding the information~\cite{tewksbury2000differences}. It may even boost recall significantly~\cite{pipps2009information}.
Building on a similar idea, we design a comparative interface where pairs of articles are displayed side by side to facilitate the comparison process for users.

\subsection{Annotating Using the Crowd and its Effect}
In educational settings, annotation has long been used to boost reading comprehension, critical thinking and meta-cognitive skill improvement.
Many online annotation tools have been developed over the last decade, including Gibeo~\cite{bateman2006oats} and HyLighter~\cite{lebow2005hylighter}.
Much of the research on annotation focuses on effects in classroom settings and often takes the form of collaborative annotation~\cite{Novak2012}.
On annotation tasks, prior research also suggests that users' performance may vary with demographic characteristics, political biases, task complexity, and subject matter~\cite{Babakar_2018,Mitra_Gilbert_2015,Hassan:2017:TAS,metzger2015cognitive,bhuiyan2020investigating}.
Some research indicates that annotation technology could improve users' effectiveness and efficiency in information-related tasks, such as search tasks~\cite{kawase2009comparison}.
Informed by such outcomes, we explore the effectiveness of comparative annotation in identifying content quality in a news consumption setting.

%% file: tex/4method.tex



\section{FORMATIVE STUDY FOR DESIGNING {\textit{NEWSCOMP}}: THINK-ALOUD INTERVIEWS}
To design an interface where users can simultaneously compare news articles, we began with a set of think-aloud interviews~\footnote{\addition{All of our studies were approved by the institutional review board at our university}}.
We used two types of prototypes during this phase: a high-fidelity interface powered by algorithmic similarity metrics and a Google Drawings board.
\addition{By high-fidelity, we mean an interactive prototype with a working front-/back-end.}
During this phase, we conducted a total of 10 think-aloud interviews with four members of our research groups (none of whom are authors of this paper) and six undergraduate students with different majors (communication, political science, and computer science) and levels of news consumption expertise.
\addition{We used a separate set of participants for interviews with each prototype.}
These interviews revealed several aspects for consideration in our design. 
Below, we discuss how our interview process evolved and summarize the insights we gathered.

\subsection{Inaccurate Algorithmic Annotation}
This phase consisted of six interviews conducted with an interface we designed to support comparison through similarity scores obtained from a state-of-the-art sentence transformer and its semantic sentence matcher\footnote{\url{https://huggingface.co/sentence-transformers/all-mpnet-base-v2}}.
All six participants mentioned that the algorithm's similarity annotations were inaccurate. 
This effect may result from differences in how users and algorithms compute similarities.
Whereas a human can take a statement, event, surrounding sentences and other contextual aspects into consideration while finding similarity, algorithms are likely to prioritize word similarity.

\begin{figure*}[t]
    \centering
    \includegraphics[width=\textwidth]{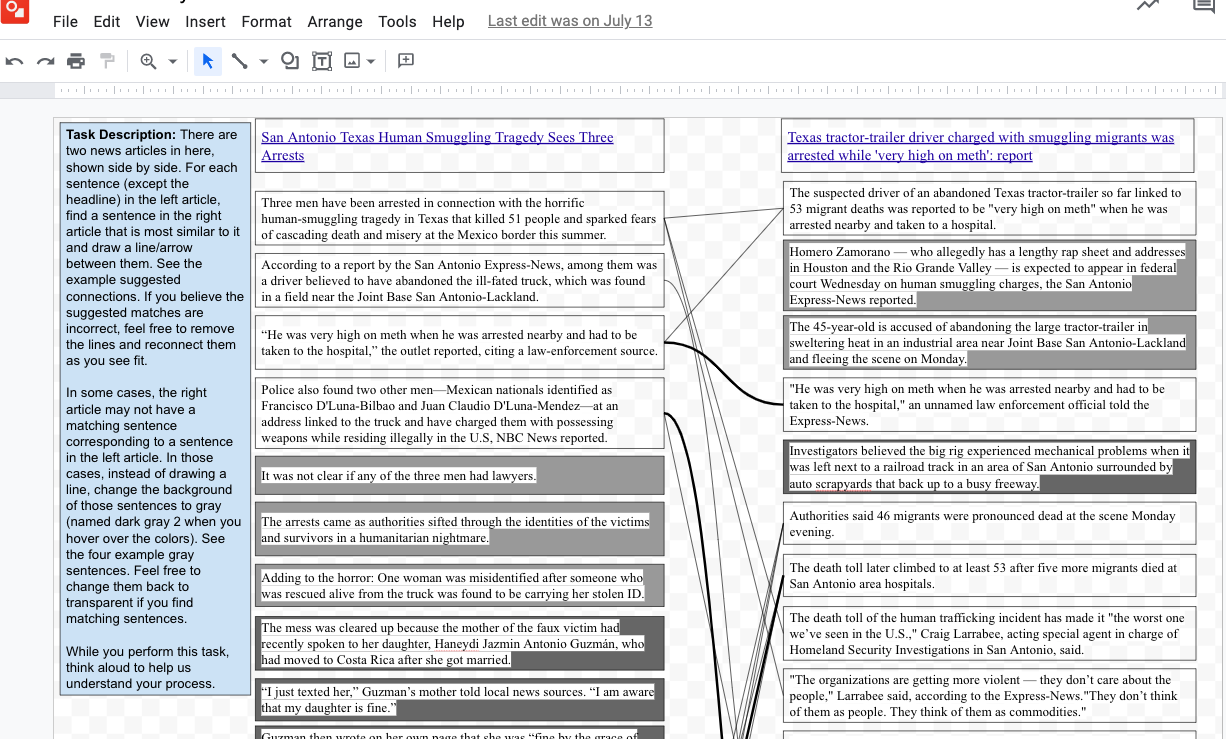}
    \caption{A Google Drawings board used for think-aloud interviews. Similar to the high-fidelity interface, two articles are presented side by side here. Users can use all the available tools to link similar statements or highlight dissimilar statements that contain important information which should be included in the other article.}
    \label{fig:tagoogle}
    \Description{This figure contains a screenshot of A Google Drawings board. Here, two articles are laid side by side with sentences separated into difference text boxes. The title of the articles contain links to the original source. Some of the sentences are connected using either straight or curved lines. Background of some of the sentences are grayed with two types of gray, dark or light gray. There is a text box on the left of both articles with instruction for the task.}
\vspace*{-9pt}
\end{figure*}

\vspace*{-9pt}
\subsection{Annotating on Google Drawings}
Next, we moved towards asking users to perform the annotation task using a Google Drawings board.
Here, we laid out a pair of news articles side by side on the drawing board by segmenting them into sentences (see Figure~\ref{fig:tagoogle}).
Then, we asked our participants to perform two annotation tasks, one after another: (i) find similar statements between the two articles and draw lines between them, and (ii) revisit statements without corresponding, similar statements to see if they convey important information that should be included in the other article.
\addition{Since users reported inaccuracies in algorithmic statement matching, we refrained from providing machine-generated annotations as suggestions.}
Figure~\ref{fig:tagoogle} shows a screenshot of the interface and the annotations provided by one of the participants.
As in the prior interviews, we recorded participants' actions. 
These tasks took longer compared to the previous interviews, as users iterated over each statement multiple times to add annotations.
After completing the task, we asked semi-structured interview questions to clarify the participants' actions.
Our observations and the participants' answers helped us identify several considerations for our study, outlined below.

\vspace*{-4pt} 
\subsubsection{Criteria for Finding Similarities: Content and Underlying Theme}
After the annotation task sessions, we asked participants to elaborate on the criteria they used to find similarities. 
From their responses, we found two similarity criteria: content and underlying theme. 
Though one participant mentioned structural position (e.g., the lede in a news article) as a criterion, none of the other participants mentioned it. 
In our final deployment, we asked users to provide rationales for why statements were similar.

\subsubsection{Considerations for Finding Important Information Present in Only One Article}
When asked about how participants chose statements conveying important information that was worthy of inclusion in the other article, they mentioned two considerations. 
The first of these considerations was whether a statement fit the narrative of the other article. Participants suggested that a statement in article A should  only be labeled ``important'' if it fits the narrative in article B and provides important context missing from article B. Such missing information might include statements detailing what happened after an event or how something happened. 
Even when a statement did not fit within the narrative of the other article, some suggested that such a statement should still be included (\inlineqt{P3}{This task is difficult, because the two articles are focusing on different narratives ... the other article does a bad job at portraying them as such [smuggled people being seen as inhuman]. Therefore, I think that bringing the human cost displayed here into the other article would be helpful.}). 
Participants mentioned that any information among the dissimilar statements that could be inferred from other statements or the context of an article did not need to be included.
In light of these nuances in the reasoning behind answering the questions, the result suggests that participants were more critically engaged in reading and comparing the two articles during this exercise than they were during the exercise that presented ML-recommended results. 
Therefore, in our final design, we asked users to provide rationales behind annotations when finding something important to be included in the other article.

\subsubsection{Readers Have Varying Expertise in Identifying Similarities}
During these interviews, we noticed that readers' different levels of expertise in reading news and knowledge on the topic led to different annotations. 
During two interviews with participants who had less news expertise, we presented another participant's thematic connection annotations and asked if the interviewees could understand the original annotator's intentions.
Neither participant was able to explain the annotator's intentions.
These findings led us to one of our research questions; specifically, how user characteristics relating to news expertise affect comparative annotation.
 
Overall, both studies revealed to us that a comparative annotation task could strengthen users' engagement with news content, and we implemented such a task in our final design for \sys{}.
 
\begin{figure*}
    \centering
    \includegraphics[width=1\textwidth]{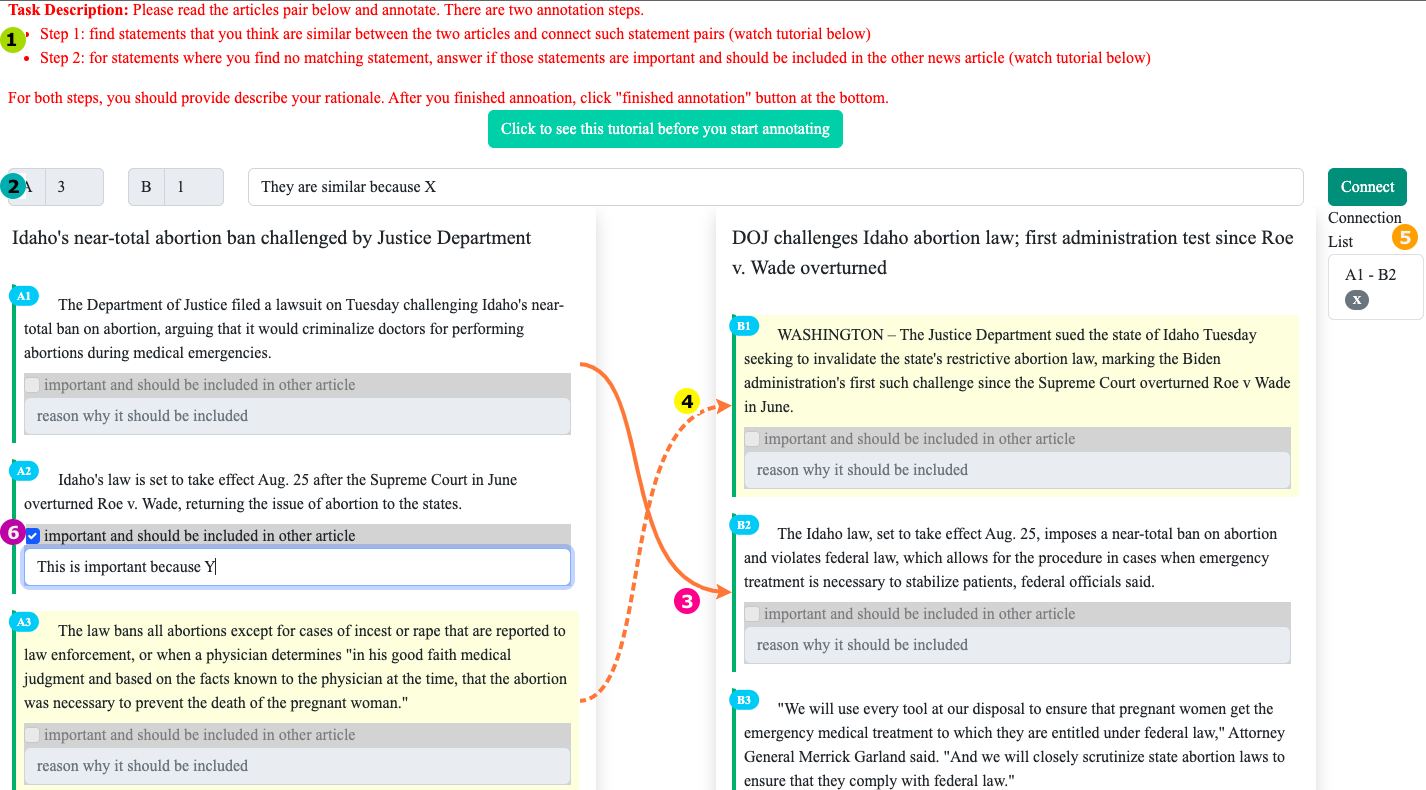}
    \caption{\sys{} Interface showcasing features with random annotations. {\includegraphics[scale=0.3]{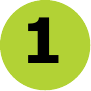}} 
 Annotation instructions in two steps: find and connect similar statements, and answer if a statement with no corresponding, similar statement is important to include in the other article. 
 {\includegraphics[scale=0.3]{figures/ot-interface-2.pdf}} 
 Toolbar to finalize a connection by providing a rationale  {\includegraphics[scale=0.3]{figures/ot-interface-3.pdf}} 
 A solid arrow representing a connection already created {\includegraphics[scale=0.3]{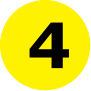}} 
 A dashed arrow indicating that the connection creation tool is active 
 {\includegraphics[scale=0.3]{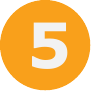}} 
 A list of connections including deletion buttons 
 {\includegraphics[scale=0.3]{figures/ot-interface-6.pdf}} 
 The importance question in step 2.}
    \label{fig:newscomp-interface}
    \Description{This figure contains a screenshot of \sys{} interface. At the top there are instructions for the task. Right below, there is a button. Below that there is a line with two selection box, a text box and a button. Below this line, there are two news articles side by side with sentences separated. Each sentence contains a selection button to identify it as important. First sentence on the left article is connected to the second sentence on the right article with a solid arrow. Third sentence on the left article is connected to the first sentence on the right article with a solid arrow. There is a text box in the right of both articles with a cross button.}
\end{figure*}


\subsection{The \sys{} Interface and How It Works}
Based on the findings from our two formative studies, Figure \ref{fig:newscomp-interface} shows the final \sys{} interface we implemented.
At the top of the page, instructions for the tasks are laid out ({\includegraphics[scale=0.3]{figures/ot-interface-1.pdf}}
). The task asks the user to read the pair of articles and perform two steps: (i) find similar statements within the articles and create links between them, and (ii) check if a statement with no corresponding similar statement in one article is important and worthy of inclusion in the other article.
To help users understand how to perform the two steps, there is a button below the task description that opens a video/GIF showing a tutorial of both steps. In our deployment, we made a point of reminding users to watch the tutorial before proceeding.
A toolbar below the taks description 
({\includegraphics[scale=0.3]{figures/ot-interface-2.pdf}}
) helps users perform the first step.
Specifically, when users link two statements, the toolbar shows the statements that are highlighted (in yellow) and provides a text box where they can supply a rationale for the connection right before finalizing the annotation.
Below the toolbar, two news articles are presented side by side (as in our initial interface) with the article title at the top, followed by statements segmented exactly as in the original article.
To limit preexisting biases, there are no links to the canonical source, nor any reference to the authorship.
The interface also hides any nontextual components (i.e., videos and images).
To begin connecting statements, users click the two statements to select them.
Selected statements are highlighted with a yellow background.
To mark a statement as worthy of inclusion in the other article, each statement contains a checkbox ({\includegraphics[scale=0.3]{figures/ot-interface-6.pdf}}
). 
There is also a text box below this checkbox for users to provide a rationale for marking a statement as important.
When a user connects two statements, the checkbox and text box for the statement are programatically disabled and grayed out.
After selecting a statement from each article, a dashed arrow representing a pending connection appears ({\includegraphics[scale=0.3]{figures/ot-interface-4.pdf}}). 
 When a user finalizes a connection by filling in a rationale and clicking ``connect,'' the dashed arrow ({\includegraphics[scale=0.03]{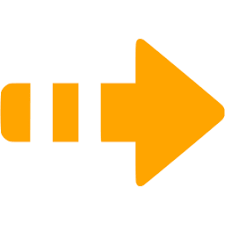}}
)  changes to a solid arrow 
({\includegraphics[scale=0.03]{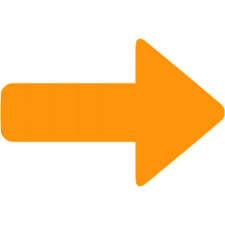}}
) 
to represent a confirmed connection ({\includegraphics[scale=0.3]{figures/ot-interface-3.pdf}}
).
In addition, a list of connections appears to the right of the articles ({\includegraphics[scale=0.3]{figures/ot-interface-5.pdf}}
) to allow users to delete connections they have created. 
Users can delete a connection by clicking the cross button next to it.
After finishing both tasks, users scroll to the bottom of the page and click a button to confirm they have finished the annotation tasks.
This system was built with a React front end \additionminor{with Bootstrap CSS}, and a Flask back-end server with a MySQL database.
To draw the connection lines, we used the Leader-Line\footnote{\url{https://github.com/anseki/leader-line}}.
To scrape news articles, we used NewsPaper\footnote{\url{https://newspaper.readthedocs.io}}.
\additionminor{For the formative study, we used Sentence Transformer\footnote{\url{https://huggingface.co/sentence-transformers}} to generate sentence embedding and calculate sentence pair similarity score.}

\begin{figure*}
    \centering
    \includegraphics[width=.8\textwidth]{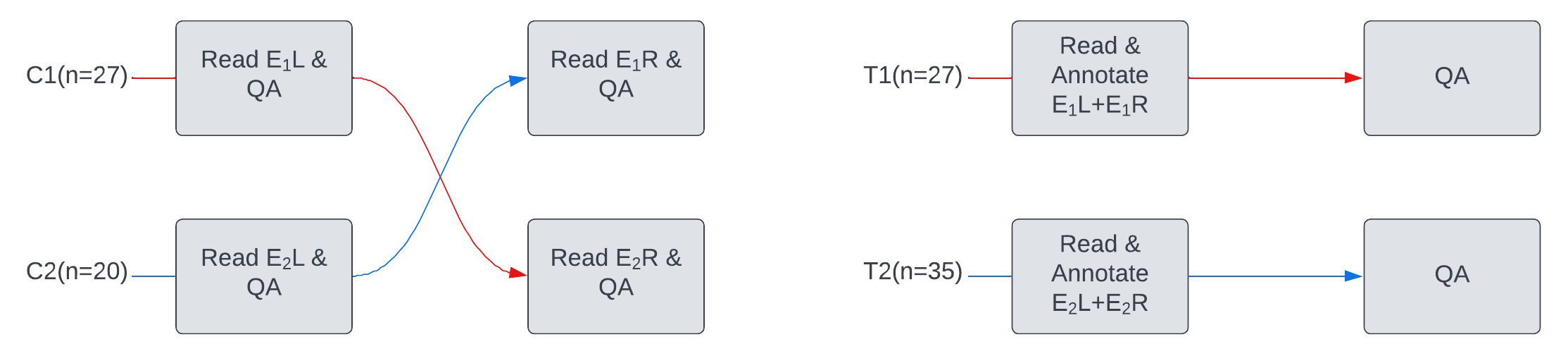}
    \caption{Study design showing the experimental conditions for each of the four participant groups. Here, C and T respectively represent control and treatment groups; the number of participants is given in parentheses. Because we used four articles, we had two control groups (C1--2) and two treatment groups (T1--2). Article E$_X$P represents an article about event X from a source with political leaning P (L for left, R for Right). Articles with $X = 1$ were about immigration, while those with $X = 2$ were about abortion. For example, E$_2$R indicates a news article about abortion from a right-leaning source. The E$_1$ pair had high contrast, while the E$_2$ pair had low contrast. In the study, we randomized the order/position of the articles for each participant. }
    \label{fig:study}
    \Description{This figure contain two plots. Each plot has four text boxes in four quadrants. The right plot contains a straight lines going through the top two boxes and another through the bottom boxes. In the left plot, the lines go diagonally.}
\end{figure*}

\section{Evaluation Study}

Using \sys{}, we examine two research questions:

\rqt{1}{\rqone{}}

\rqt{2}{\rqtwo{}}

To answer, we conducted a between-subjects experiment in a controlled environment using two pairs of news articles.
We created a separate interface for the  control users. 
In the control interface, only one article is shown at a time.
Figure~\ref{fig:study} shows the study design for our experiment with four experimental groups: two treatment and two control groups.
Each group read two news articles. While the treatment group was able to view two articles on the same topic simultaneously, the control users read articles on different topics sequentially to account for any learning effects from recall and comparison.
All four groups read stories from two sources with different political leanings.
We randomized article location (left or right) for the treatment group and article order (first or second) for the control group to account for any ordering effect.

\vspace*{-9pt}
\subsection{Article Selection}
For our study, we picked two politically contentious topics (immigration and abortion), where reading content from diverse perspectives can be beneficial. The topics were chosen from recent news coverage at the time of the user studies. 
For each topic, we chose articles published at least two weeks prior to deployment to limit possible recall effects. 
Pairs were selected by finding two articles from politically opposed sources under the same story bundle on Google News. 
When choosing article pairs, we picked pairs with different levels of similarity and difference.
Since the article pair on abortion(E$_2$) had more similarities than differences, we categorized the pair into the \textit{low-contrast} category.
On the other hand, the pair on immigration (E$_1$) had more apparent differences than similarities, so we categorized it into the \textit{high-contrast} category.
This categorization was confirmed by our experts' gold standard annotations (see \ref{sec:goldstandard}), which identified more than 50\% of the article text as similar in the low-contrast pair while identifying less than 25\% of the text as similar in the high-contrast pair.
The selected articles (E$_1$L, E$_1$R, E$_2$L, E$_2$R) are reproduced in Appendix~\ref{articles}. 

\begin{table*}
    \centering
    \small
   \begin{tabular}{p{0.10\textwidth}p{0.85\textwidth}}
   \hline
   Credibility~\cite{meyer1988defining} & \begin{tabular}{p{0.85\textwidth}}
        i) It is biased (I)\\
        ii) It is not fair (I)\\
        iii) It doesn't tell the whole story (I)\\
        iv) It is not accurate (I)\\
        v) It cannot be trusted (I)\\
    \end{tabular}\\
    \hline
    Quality~\cite{urban2014news} & \begin{tabular}{p{0.85\textwidth}}
        i) It shows multiple viewpoints\\
        ii) It has information on causes and consequences\\
        iii) It provides balanced viewpoints\\
    \end{tabular}\\
    \hline
   Current Event Knowledge (CEK)~\cite{maksl2015measuring} & \begin{tabular}{p{0.85\textwidth}}
        i) Who is/was Kamala Harris? (a) President (b) \textbf{Vice President} (c) \textbf{Senator from California} (d) UN Ambassador\\
        ii) What does the recent Supreme Court ruling overturning Roe v. Wade entail? (a) \textbf{Abortion is not a constitutionally protected right} (b) In Missouri, abortion is legal before 24 weeks (c) All US states allow abortion for rape and incest (d) \textbf{There is confusion about abortion rights relating to miscarriage and ectopic pregnancy}\\
        iii) In California (a) \textbf{everyone, including undocumented individuals, has the right to access their crime report} (b) there are no immigrant detention facilities (c) \textbf{state and local police officers cannot inquire about an individual’s immigration status during a routine check} \\
        iv) How is the Fed responding to the high inflationary economic condition? (a) \textbf{Raising the interest rate} (b) Lowering the interest rate (c) Keeping the interest rate the same\\
   \end{tabular}\\
   \hline
   Value of Media Literacy (VML)~\cite{vraga2015multi} & \begin{tabular}{p{0.85\textwidth}}
         i) Two people might see the same news story and get different information from it\\
         ii) People are influenced by news whether they realize it or not\\
         iii) News is designed to attract an audience's attention\\
         iv)Writing techniques can be used to influence a viewer's perception\\
         v) People should accept information from the news on face value (I)\\
         vi) It is the job of citizens to overcome their own biases in consuming news\\
         vii) People need to critically engage with news content\\
         viii) The main purpose of the news should be to entertain viewers (I)\\
    \end{tabular}\\
    \hline
   \end{tabular}
   \caption{Questionnaires used in the study. Credibility and quality questions were asked after reading or annotating. (I) means these items were inverted for analysis. The correct responses appear in boldface. The CEK questionnaire contains multiple-choice questions, while the VML, credibility, and quality questions are 5-point Likert items. The VML and CEK items were presented in the pre-survey. }\label{tab:newscomp-qa}
\end{table*}

\subsection{Measuring Credibility, Quality, Current Event Knowledge, Media Literacy}
To address RQ1, we measured two expertise metrics: \textit{current event knowledge} and \textit{value of media literacy}.
Here, the value of media literacy differentiates users' general media literacy from their expertise on topics related to our study.
To answer RQ2, we use perceptions of article \textit{credibility} and \textit{quality}, and we compare the treatment groups' assessments with the control groups'.
Below, we discuss how we measured each metric.

\subsubsection{Current Event Knowledge (CEK) and Value of Media Literacy (VML)}
To capture users' news-related knowledge, we adapted the Current Event Knowledge measure created by Maksl et. al.~\cite{maksl2015measuring}.
Here, we included questions relevant to the two chosen article topics and some other timely topics (see Table~\ref{tab:newscomp-qa}).
To measure users' perceptions of media literacy, we used a prior scale created by Vraga et. al.~\cite{vraga2015multi}.
To calculate CEK scores, we added 1 point for each right answer and deducted 1 point for each wrong answer.
In our study, CEK ranged from -1 to 7, with 4 being the median value.
For VML, we average the responses across items.
The score for VML ranged from 1 to 8, with 6 being the median.
Finally, for both measurements, we use the median score to create a binary response variable with values ``low'' and ``high.'' For example, users scoring less than 4 in CEK were categorized as low-CEK users and vice versa.

\subsubsection{Credibility \& Quality}
We used a five-item questionnaire by Meyer et al.~\cite{meyer1988defining} to measure users' perceptions of credibility for every news item (see Table~\ref{tab:newscomp-qa}).
In our study, we found that this measure had high internal consistency (Cronbach's $\alpha = 0.85$), close to the result in Meyer et al.
For news quality detection, we use a modified version of the questionnaire suggested by Urban et al.~\cite{urban2014news} (see Table \ref{tab:newscomp-qa}).
Similarly to the credibility questionnaire, participants' responses to these questions showed high internal consistency (Cronbach's $\alpha = 0.89$).
We measured all of these items on a 5-point Likert scale, from ``Strongly Disagree'' (1) to ``Strongly Agree'' (5).
Note that scores for the credibility items are inverted for analysis.

\subsection{Recruitment}
To recruit participants for our final \sys{} interface, we used Facebook advertising for two weeks in August 2022.
This method allowed us to organically recruit diverse participants from a large pool.
We also did limited advertising on news subreddits (such as, \texttt{r/politics, r/moderatepolitics, r/news, r/neoliberal, and r/conservative}) through private messages from our research group's Reddit account, reaching about 40 users.
Two users responded to these messages.
Since the article topics are US-centric, our ads targeted people living in the US with interest in news-related pages.
Thus, our study result may not be generalizable beyond the context of the US.
The advertisement led users to a pre-survey to sign up for the study.
In the pre-survey, we screened users with the following study eligibility criteria: (i) I am 18 years old or over, (ii) I reside in the United States, (iii) I read at least one news article online every day, (iv) My primary language for news consumption is English, and (v) I use a laptop or a desktop for online news reading.
Besides these criteria, we also screened out users who failed attention checks, had an IP address outside of the US, or spent very little time (less than half of the median time, which was two minutes) in the pre-survey.
Overall, 685 users clicked on the survey, out of which 238 passed the screening criteria.
We invited all of these participants to the study in multiple batches.
Ultimately, \participants{} participants completed the study.
Participants who completed the study were compensated with \$7.50 gift cards for the 30-minute study, in line with the state's minimum wage.
 
\subsection{Procedure}
Users who met the screening criteria in the pre-survey filled out the rest of the survey, which contained questions about demography, including gender, age, race, education, and political affiliation, and the two news expertise measures.
Within three days of submitting the survey, we invited eligible users to participate in the study via email.
In the email, we provided the consent document, instructions for using the interface, and a link to the study website.
When users clicked the link to access the study website, they were randomly assigned to one of four groups (two treatment and two control) to ensure a balanced sampling design.
Since some people who clicked the link ultimately did not complete the study, the final group sizes are not exactly equal.
Recall that after visiting the study website, treatment users were asked to view a tutorial on how to add annotations before reading and annotating articles.
After finishing the annotation process, users responded to the credibility and quality questionnaire.
The annotation interface, including the articles and annotations, was still visible at this time.
In the control condition, there was no annotation task, and participants read only one article at a time.
The control users additionally responded to the credibility and quality questionnaire after reading each article (see Figure~\ref{fig:study}).


\begin{figure*}
    \centering
    \includegraphics[width=1\textwidth]{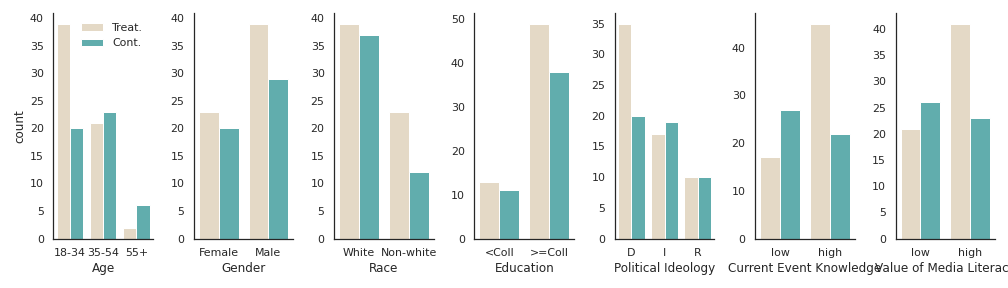}
    \caption{Graphs showing the distribution of participant demographics across the treatment and control groups.}
    \label{fig:demo}
    \Description{There are seven bar plots in this figure showing bars for treatment and control users separated by seven demographic criteria.}
\end{figure*}

\subsection{Participant Pool}
Due to screening and self-selection bias, our study participants were not equally distributed in certain demographic dimensions, such as age.
Additionally, since one of our aims was to identify how users with different demographic characteristics compare in their annotations in RQ2, we invited more users in the treatment condition.
Though our pre-survey had a large number of categories for different demographic characteristics, we merged groups with small numbers of respondents for more meaningful differentiation.
Figure~\ref{fig:demo} shows this distribution.
Here, we grouped two consecutive age groups, merged participants from nonwhite races together, and divided respondents by education into those with any university degree versus those with no degree.
Generally, participants were skewed towards younger age groups, male, white, college-educated, and politically left-leaning.

\subsection{Gold Standard Generation}
\label{sec:goldstandard}
To compare annotation quality, we used expert-produced gold standards.
To obtain gold standards for both the annotations and the perception metrics, we recruited two senior PhD students from the university's Department of Communication for an interview session. 
Both had past experience in conducting news content analysis research and were familiar with both topics used in our study.
One of the experts also worked as a journalist for more than five years.
To generate credibility and quality perception scores, both were given the original links to the news articles and asked to rate the RQ1 questions\footnote{This task was performed before generating annotations. We did not ask them to work within the comparative interface, with the assumption that they would rate the articles accurately irrespective of any comparison.}.
They were also allowed to do any outside research they wished.
After rating their perceptions, we provided the article pairs in a Google Drawings board and asked them to add annotations, much like the process from our think-aloud interviews.
After adding annotations independently, the expert annotators met with each other to resolve any conflicts.
Through this method, we built consensus around our gold standard annotations.


%% file: tex/5result.tex
\section{Results}
To answer both research questions, we compared users' annotation and perception responses against the expert-produced gold standards. 
For this purpose, we performed a series of analyses involving mean testing, analysis of variance, and regression.
For free-form text responses (specifically, the annotation rationales), three authors performed thematic coding (\addition{see supplemental document for data and codes}).
Below, we outline the results.
 
\subsection{RQ1: \rqone{}}

\begin{figure*}
    \centering
    \includegraphics[width=0.9\textwidth]{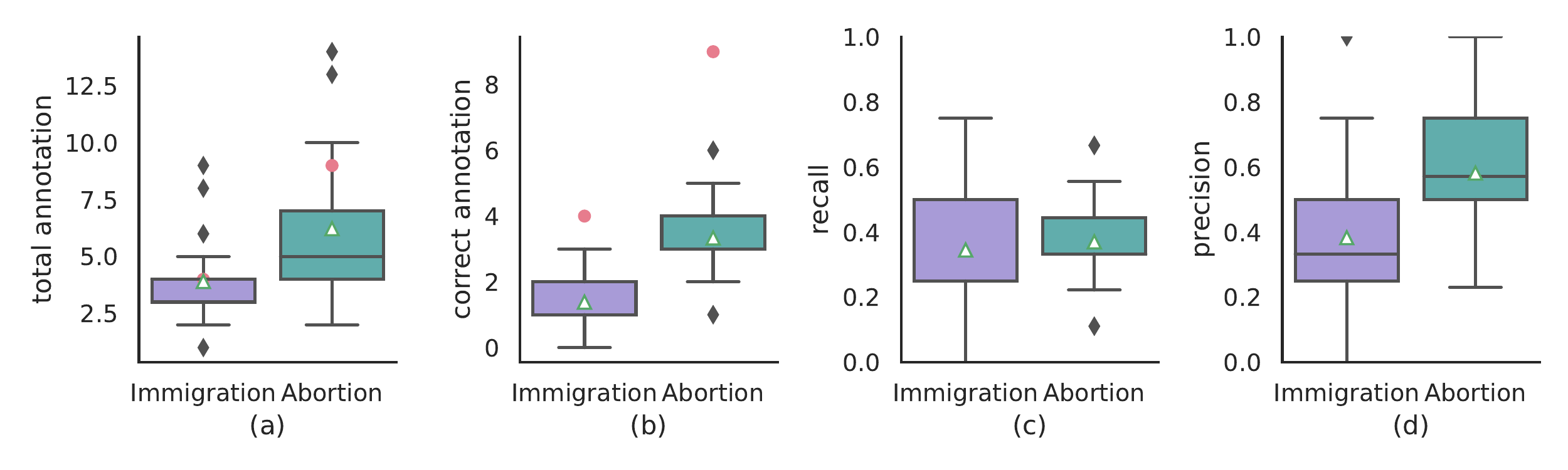}
    \caption{Distribution of connection making by users. White and red dots respectively represent the average and experts' annotation.}
    \label{fig:conn}
    \Description{There are four box plots in this figure showing boxes for each topic (immigration and abortion). The four plots are respectively from left to right for total number of annotation by users, total correct number of annotation by users, precision and recall.}
    \vspace{-8pt}
\end{figure*}

\subsubsection{Performance on Connection-Making}
Figure \ref{fig:conn} shows the distribution of total connections users made, correct connections made, their recall, and precision relative to the gold standard.
The median number of connections between articles fell below the gold standard for both article pairs, as shown in Figure~\ref{fig:conn}(a).
Between the two article topics (immigration and abortion), users on average made more connections---correct or otherwise---between the abortion articles (the low-contrast article pair).
Furthermore, users' precision was significantly better on the abortion articles than on the immigration articles (\addition{Mann-Whitney U = 136.5,} p < 0.001).
However, we did not find any significant difference in recall.
 
\begin{figure*}
    \centering
    \includegraphics[width=1\textwidth]{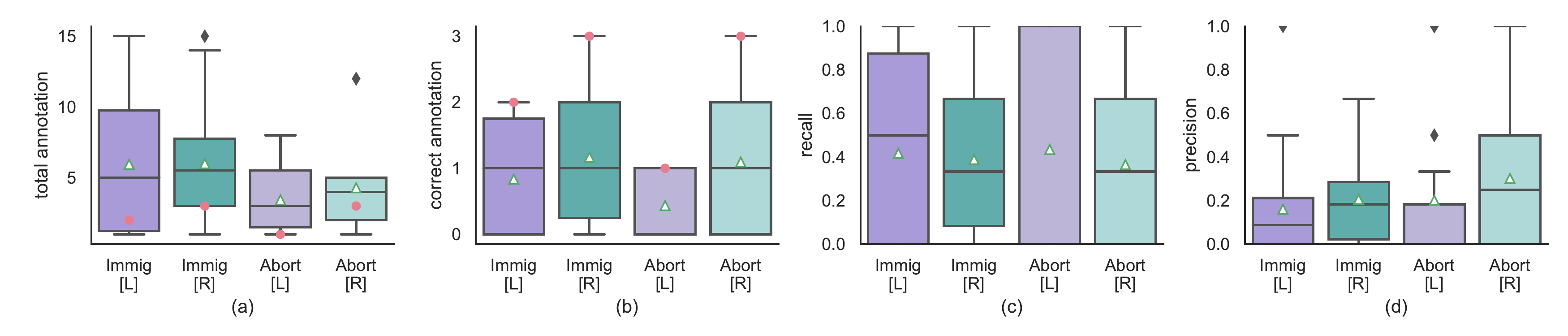}
    \caption{Distribution of importance detection by users. White and red dots respectively represent the average and experts' annotation.}
    \label{fig:imp}
    \Description{There are four box plots in this figure showing four boxes each by topic (immigration and abortion) and the leaning of those article sources (left and right). The four plots are respectively from left to right for total number of annotation by users, total correct number of annotation by users, precision and recall.}
\end{figure*}

\subsubsection{Performance on Importance Detection}
Figure~\ref{fig:imp} shows the distribution of total importance annotations users made, correct importance annotations, recall, and precision relative to the gold standard.
As shown in Figure~\ref{fig:imp}(a), the median number of importance annotations was consistently above the gold standard.
Between the two article pairs, users on average annotated more items as important---correctly or otherwise---in the immigration articles (the high-contrast article pair).
Though users' recall was high due to the large numbers of importance annotations added, their precision was low, with the median per article being less than or equal to 0.25.
Comparatively, for connection-making annotation task, users' median precision and recall are higher than these median for importance annotations.

\begin{figure*}
    \centering
    \includegraphics[width=0.95\textwidth]{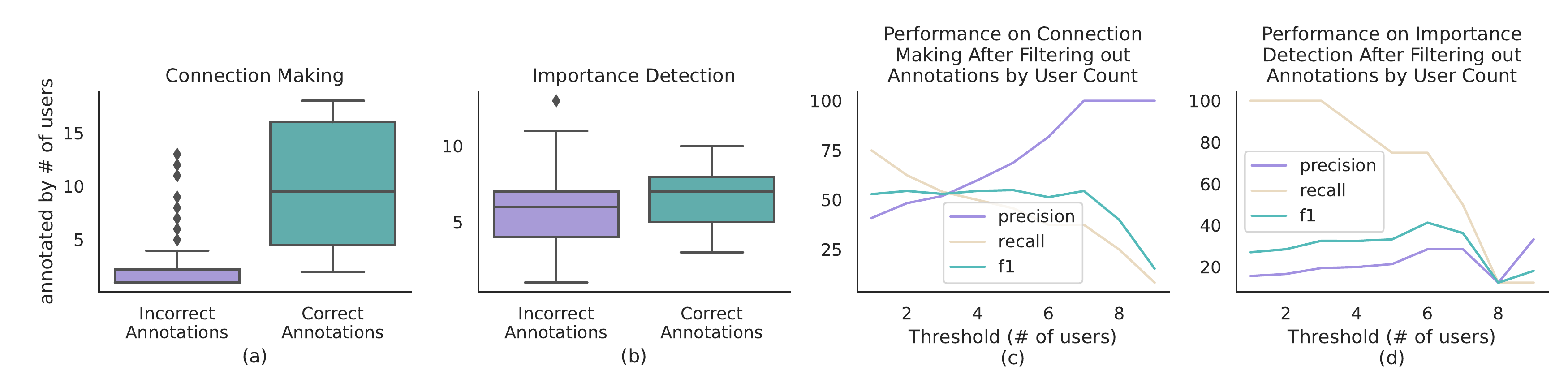}
    \caption{(a \& b) User agreements on incorrect and correct annotations. (c \& d) We filtered annotations by the number of concurring users to see how annotation performance changes as the threshold moves. Here, for connection making and importance detection, the F1 scores peak at five (55\%) and six (41\%) users, respectively. }
    \label{fig:ann-agree}
    \Description{There are two box plots on this left side of this figure and two line plots on the right. The two left plots shows boxes for correct and incorrect annotations for two annotation tasks. Each plot on the right contains three lines (precision, recall and F1 score) by varying the threshold to use to filter out annotations for each task.}
\end{figure*}
\subsubsection{Annotation Agreement}
Next, we examined how users agreed on annotations among themselves by plotting the count of users annotating each item.
Figure~\ref{fig:ann-agree} shows the distribution of this analysis.
Here, we differentiated between agreement on correct and incorrect annotations.
For the connection-making task (Figure~\ref{fig:ann-agree}(a)), we found that the annotation count for correct items was significantly higher than the count for incorrect items (\addition{Mann-Whitney U = 517.0,} p < 0.01).
However, for importance detection (Figure~\ref{fig:ann-agree}(b)), the corresponding counts did not differ significantly.
Furthermore, we also observed some outliers (high agreement in some annotations) in the connection-making annotation task not made by the experts.
In Figures~\ref{fig:ann-agree}(c) and (d), we examine how annotation performance changes by filtering annotations by the number of concurring users.
Overall, a threshold of five users produces the highest F1 score (55\%) for the connection task, while peak performance (41\%) occurs at a threshold of six users for the importance detection task.

\begin{figure*}
    \centering
    \includegraphics[width=0.85\textwidth]{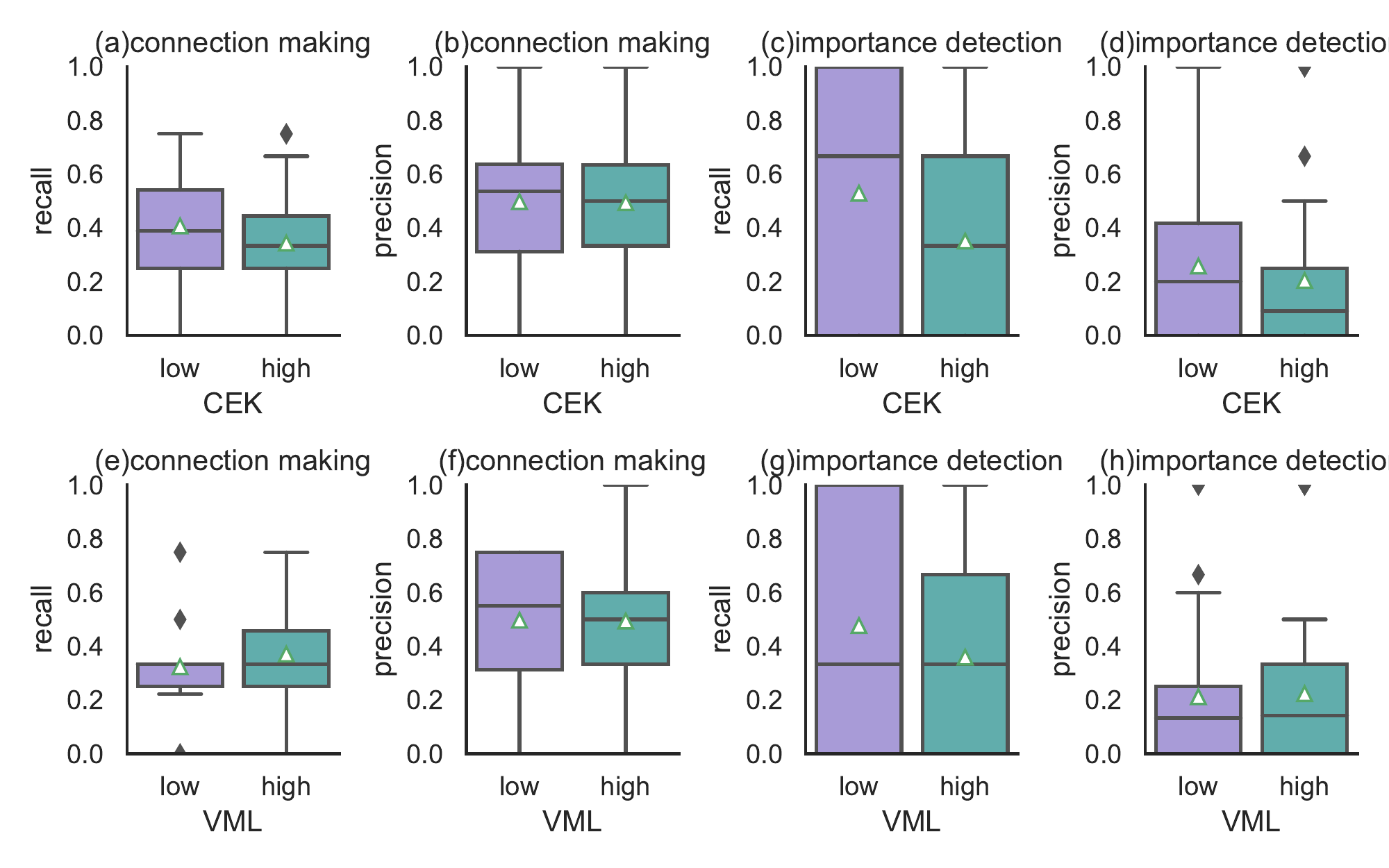}
    \caption{Distribution of recall and precision for connection-making and importance detection divided into low/high CEK users (top), and low/high VML users (bottom).}
    \label{fig:news-exp}
    \Description{There are eight box plots in this figure with four boxes on the left for connection making task and four on the right for importance detection task. The four on the top are for recall and precision divided by low and high CEK users where the bottom ones are divided by low and high VML.}
\end{figure*}

\subsubsection{\addition{Effect of News Expertise}}
\addition{We investigated whether levels of news expertise affect users' annotation performance with \sys{} by performing Mann-Whitney U tests on precision and recall scores (for connection-making and importance detection).
Figure~\ref{fig:news-exp} shows the distribution of those scores divided by two news expertise criteria, Current Event Knowledge (CEK) and Value of Media Literacy (VML) perception.
Here, only for recall scores on the importance detection task (Figure~\ref{fig:news-exp} (c)), we found significant differences in values between low and high CEK (Mann-Whitney U = 810.5, $p<0.05$).
None of the other tests detected a statistically significant difference. Furthermore, we modeled these variables against user characteristics with a series of linear models (M1--M4 in Table~\ref{tab:ann-user} in Appendix \ref{app:user-char}). These models were similarly significant ($\beta=0.35, p<0.01$ in M1). However, since the model effect sizes ($R^2$) are low (\~0.10), there may be confounding variables not accounted for in these models affecting the outcome. It is therefore difficult to make any strong claims in this regard, and we instead leave this to future experiments.}
   
\begin{table*}
    \centering
    \small
    \begin{tabular}{p{0.08\textwidth}p{0.12\textwidth}p{0.40\textwidth}p{0.35\textwidth}}
    & \textbf{Code} & \textbf{Definition} & \textbf{Example Response}\\
    \hline
         \multirow{12}{*}{Connection} & Empty (20\%) & Empty or N/A response \\
         & Person (19\%) & Mentions that both statements refer to one or more persons involved in an event (not including quotes) & \textit{They are similar because they both mention \textbf{the owners of the truck}} \\
         & Location (1\%) & Mentions that both statements refer to the same location where an event occurred & \textit{This excerpt shows where the truck was found and both gave \textbf{an identical location} ...} \\
         & Date (2\%) & Mentions that both statements refer to a single date when an event occurred or will occur & \textit{Both statements note that the ban will take effect on \textbf{August 25th}}\\
         & Quote (9\%) & Mentions that both statements reference either the same quote or different quotes from the same person & \textit{They are similar because both highlight \textbf{a quote from Becerra (HHS Secretary)} insisting ...} \\
         & Information (48\%) & Mentions that both statements contain the same information describing the what, why, or how of the event & \textit{Similar because [both] discuss \textbf{Medical Treatment and Labor Act}}\\
         \hline
         \multirow{9}{*}{Importance} & Empty (29\%) & Empty or N/A response & \\
         & Important (15\%) & Mentions that a statement is important without providing a reason & \textit{[because it is an] \textbf{important part} of the news} \\
         & Clarification (43\%) & Mentions that a statement clarifies or elaborates on other statements & \textit{The \textbf{statement in the other article} from Becerra (HHS Secretary) is \textbf{confusing}.}\\
         & Missing (8\%) & Mentions that a statement presents a perspective missing from the other article & \textit{\textbf{No statement from Lawrence} in the other article} \\
         & Factual (4\%) & Mentions that a statement is factual and not an opinion & \textit{\textbf{Facts here.} It isn't opinion being interjected into a news story.} \\
        \hline
    \end{tabular}
    \caption{Coding scheme for annotation rationales.}
    \label{tab:rational}
\end{table*}

\subsubsection{Reasons behind the Annotations}
\begin{figure*}
    \centering
    \includegraphics[width=0.9\textwidth]{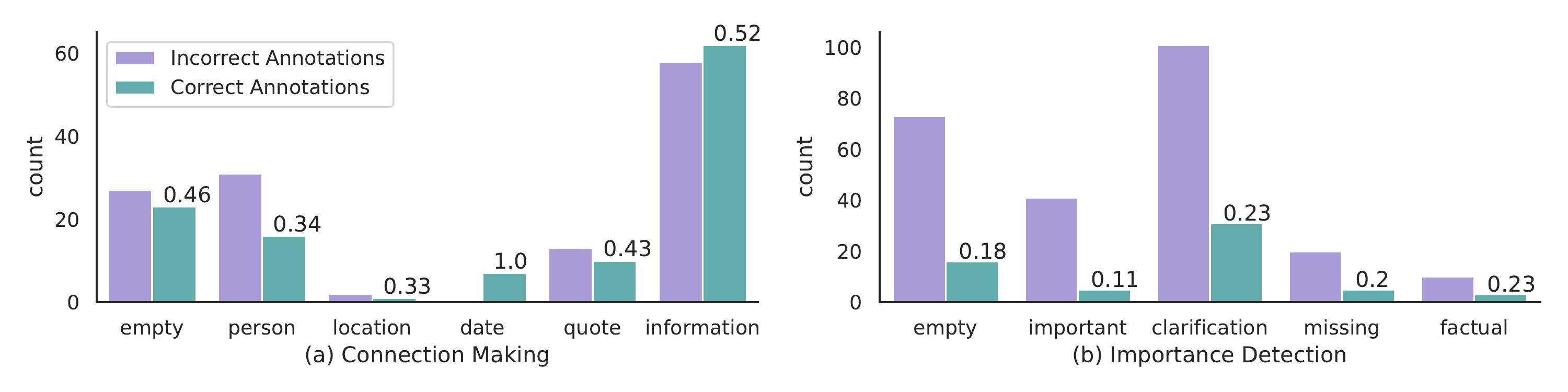}
    \caption{Annotation counts by coded rationales divided into correct and incorrect annotations. The numbers over the bars represent the ratio of correct to incorrect annotations within each code.}
    \label{fig:qual-code}
    \Description{There are two bar plots in this figure for connection making and importance detection annotation. Each plot shows counts for different rationales divided into correct and incorrect annotation.}
\end{figure*}

Three of the authors thematically coded the rationales provided by the participants during annotation.
Each author performed initial coding and discussed the results with the others to agree upon a code book.
Then, the first author coded the responses accordingly and the others checked the final codes.
Participants annotated 250 connections and 305 important statements.
Table~\ref{tab:rational} shows the coding scheme we developed for each annotation task with sample responses matching the code.
There are six codes for connection-making and five codes for importance detection (including one in each for empty responses).
Here, the codes in the connection-making task offer potential answers to the 5W1H questions (Who, What, Where, When, Why, and How).
For example, the code  ``Person'' answers the \textit{who} aspect of the event, while the code ``Information'' answers a combination of what, why, and how questions.
For importance detection, users sometimes claimed that a statement was important without explaining the reason for this assessment.
In other cases, users mentioned that a statement was important because it clarified or elaborated on existing statements, or because it provided an account from a missing perspective.
Figure~\ref{fig:qual-code} shows the count of each rationale in terms of the developed codes, divided into correct and incorrect annotations. 
For connection-making (Figure~\ref{fig:qual-code}(a)), we found that a majority of users identified similarities when the same information was presented in both articles, followed by mentions of the same person. 
For importance detection(Figure~\ref{fig:qual-code}(b)), many responses were coded into the clarification and elaboration categories, followed by the empty response category.
Figure~\ref{fig:qual-code} suggests that the rationales are not distributed proportionally for correct and incorrect annotations. 
Notably, users appear more likely to make errors in certain cases.
For example, for importance detection (Figure~\ref{fig:qual-code}(b)), the ratio of correct and incorrect ``important'' annotations shows that these annotations are more likely to be mistaken than others.
Although we performed regression on the codes to differentiate correct and incorrect annotations, the model effect sizes were very low for both models ($R^2<0.05$).

\begin{figure*}
    \centering
    \includegraphics[width=0.95\textwidth]{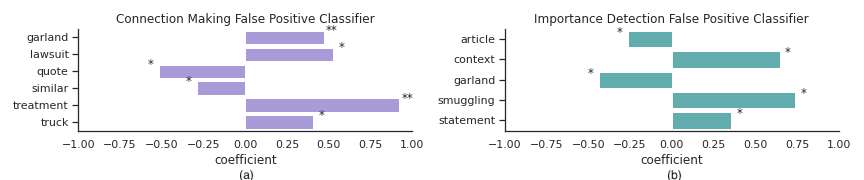}
    \caption{False positive detection with OLS using the top 50 TF-IDF words in users' responses. Here, we listed only words with significant coefficients. For example, when users' mentioned ``quote'' in a rationale, the annotation was less likely to be erroneous. On the other hand, when users mentioned the general nature of the event (``lawsuit'' in this example), the annotation was more likely to be erroneous. The model effect sizes ($R^2$) were 0.34 and 0.22, respectively.}
    \label{fig:fpr}
    \Description{There are two bar plots in this figure for connection making and importance detection annotation. Each plot shows bar of beta coefficient score for several keywords.}
\end{figure*}

Since differentiating correct and incorrect annotations by codes did not work well, we fit two models to predict incorrect annotations (false positives) for both the connection-making and importance detection tasks using the top 50 TF-IDF~\footnote{\addition{TF-IDF stands for term frequency–inverse document frequency, a statistic representing how important a word is to a document in a collection of documents}} text features from users' responses on the rationals.
Figure~\ref{fig:fpr} shows the text features with significance for this analysis.
Examining the words with significant coefficients, we can see that some words are generic (e.g., ``quote'', ``similar'', ``context''), while others are article-specific (e.g., ``Garland'' (the current attorney general), ``lawsuit'', ``smuggling'').
These differences suggest that such generic words in rationales can be used across articles to differentiate false positives from true positives, while specific words may not be usable.
We discuss these results further in section~\ref{sec-dis1}.

\begin{figure*}
    \centering
    \includegraphics[width=0.95\textwidth]{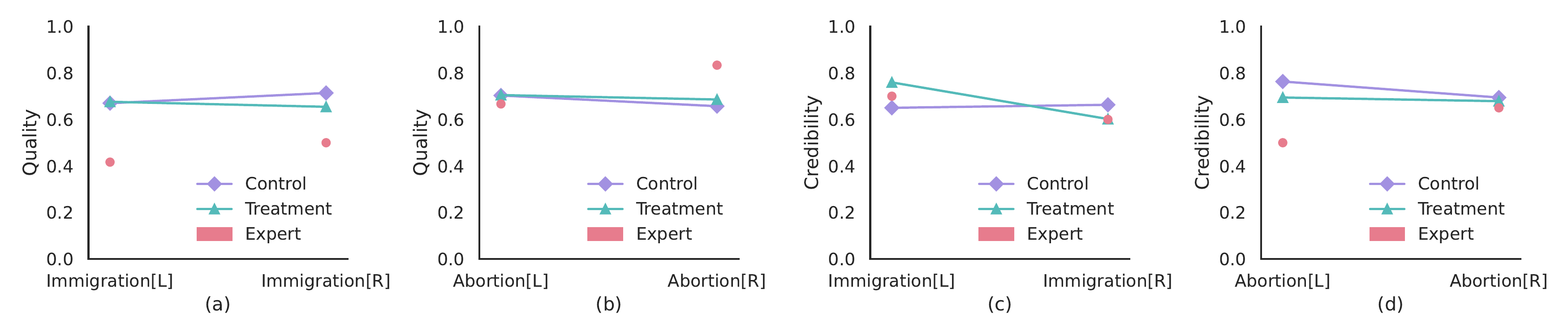}
    \caption{Interaction effects of groups and articles. We only found a marginal interaction effect (p=0.052) for credibility score on articles regarding immigration (c).}
    \label{fig:rq2}
    \Description{There are four interaction plots in this figure. Two on the left are for quality score and two on the right for credibility score. Each plot is for different article topics and contains two lines and two red dots. Lines indicate changes for treatment and control groups score. Dots indicate experts' perception.}
\end{figure*}

\subsubsection{\addition{Comparative Perception between Article Pairs}}
\addition{After the annotation task, in addition to asking about credibility and quality perception, we asked users what they noticed when comparing the two articles (``Comparing the two articles, what else did you notice about how each portrayed the issue?'').
Analyzing the responses, we found five themes, summarized in Table~\ref{tab:comp-perc}.
Notably, more than one fourth of the participants remarked on informational placement or depth (16/62), perspectives and biases (22/62) and factuality/opinions (16/22).
Some also noticed empathetic news reporting (5/62) and the use of inflammatory language (3/62).
}

\begin{table*}
    \centering
    \begin{tabular}{p{0.25\textwidth}p{0.70\textwidth}}
         \textbf{Theme(n)} & \textbf{Example Response}\\
         \hline
         Perspectives and Biases (22) & They were taking \textbf{different sides} of the equation and putting forward different thought processes\\
         Information Placement or Depth (14) & The right article was \textbf{less descriptive} and focused more on the restrictions and not the case. It was definitely telling the story from one point of view. The article on the left was very informative and unbiased\\
         Factuality or Opinions (16) & Article B provided responses from the Idaho government, whereas Article A did not include \textbf{commentary} from Idaho, but instead from Texas which did not seem relevant\\
         Empathetic Reporting (5) & There were more \textbf{humane aspect} in the article on the left\\
         Inflammatory Language (3) & Article B seemed to be making the issue out to be more \textbf{controversial} by going back and forth between perspectives more frequently\\
         \hline
    \end{tabular}
    \caption{\addition{Themes in users' responses to a question asking what they noticed about the two articles overall. Note that while an example response may belong to multiple themes, only the portion relevant to the listed theme is presented in bold.}}
    \label{tab:comp-perc}
\end{table*}

\subsubsection{\addition{Perception of the Tool}}
\addition{Apart from the task-specific questions, we also asked annotators about their perceptions of the \sys{} tool. 
Overall, perceptions of the tool were split among positive (28/62), neutral (21/62), and negative (15/62) sentiment.
The reasons behind negative sentiment included the lengthy nature of the task (3/15), difficulty in performing annotation (8/15), and confusion regarding the instructions (4/15).
While it may have been hard in the beginning, users quickly learned how to use the tool (\inlineqt{U1}{It was a bit confusing to learn how to use the tool, but it was easy to use once I played around with it.}).
Improvements to the tool design could potentially address these issues.
For instance, during connection-making, a search tool could assist users with finding similar statements quickly.
Users also suggested improvements such as allowing them to set the weight of the connected lines, change the colors of lines, and see how others annotated a statement.
}

\subsection{RQ2: \rqtwo{}}
To answer this question, we performed a two-way ANOVA on the response variables, credibility scores, and quality scores.
To calculate each score, we first summed item scores for the questionnaire on each score (credibility and quality) and standardized them on a [0, 1] interval.
Then, we performed the two-way ANOVA on the group and article interaction. 
Figure~\ref{fig:rq2} shows the result of this analysis.
We did not find any significant difference in quality perceptions.
We also performed a one-sample \textit{t}-test on users' quality perception responses against those of the experts.
Though we did find some similarity for the abortion articles (Figure~\ref{fig:rq2}(b)), in the case of the immigration articles (\ref{fig:rq2}(a)); that is, the high-contrast pair, the difference between the users' and experts' ratings was significant.
This result suggests that neither group performed well on the quality question, especially for the high-contrast article pair.
  
Next, we analyzed the interaction of credibility with the articles.
In Figure~\ref{fig:rq2}(c), our analysis found that the interaction between the experimental group and the article is marginally significant for the articles on immigration with a close to moderate effect size (F(1) = 3.83, p = 0.052, Cohen's f = 0.22).
We did not find any significant effect for articles on abortion.
\addition{Note that when we fit a mixed-effects regression model on the same data additionally considering repeated measure, we found significant interaction effects on the high-contrast articles (see Appendix~\ref{app-rq2}).}
Moreover, when we look at the articles on abortion, despite the experts not performing comparative annotation, their perceptions of one article (from a left-leaning source) were significantly lower than those of the other (from a right-leaning source).
During our discussion, we found that experts used certain criteria to arrive at this assessment~\footnote{These articles talk about the DOJ's challenge against an abortion restriction law in Idaho. Here, the experts mentioned that the article from the left-leaning source did not mention an opposing perspective, such as that of Idaho's government. Note, however, that the article did cite the state attorney general of Texas, who supports abortion restrictions.}.
On the other hand, the comparative annotation task may have given users the impression that both articles are highly similar (leading to similar ratings) and discouraged them from examining other differences too closely.
This result suggests that there is room for improvement in task designs for comparative annotation.
We discuss further implications in section~\ref{sec-dis2}. 

%% file: tex/6discussion.tex
\section{Discussion}
Through our experiment, we found that users generally perform poorly in annotation tasks for finding similar statements and identifying important statements among statements with no similarity between two news articles.
However, they have better precision in finding similarities than in identifying important statements.
We found that when a high number of users find two statements similar, such annotations have a high chance of coinciding with experts' annotations.
Furthermore, we found that users with low current event knowledge may perform better annotations.
Analyzing users' rationales behind the annotations, we found several reasons for finding similarities (e.g., mentions of the same person or information) and identifying important statements among statements with no similarity (e.g., statements clarifying something or providing a missing perspective).
Furthermore, we found that certain words have significant power in differentiating false and true positives.
\addition{After annotation, users also mentioned noticing differences in how article pairs represented things, such as perspectives, information placement, information depth, and facts/opinions.}
In RQ2, we found that annotation tasks may have limited effects on users' perception of news credibility for high-contrast news articles.
Below, we discuss the implications of these results.
 
\subsection{RQ1: Annotation Performance}\label{sec-dis1}
Our results indicate that although users perform poorly in general, their performance varies across annotation task types and article pairs, depending on the degree of contrast between the articles in the pair.
Compared to the experts, average users seem to find fewer connections and consider more statements worthy of inclusion in the other article in a pair.
This difference shows how experts and users differ when reading two news articles.
This difference could stem from analytical capabilities; that is, perhaps finding similarities and differences is a task that requires particular expertise relating to news.
Or, perhaps low knowledge users are more attentive to the articles.
Another reason behind the difference could be users' preconceived biases regarding the media in general~\cite{USMediaP92online}.
Such perceptions may have influenced users to see fewer connections and more differences.
Therefore, one direction for future research may be to examine the rationales behind identified differences by varying the task complexity and user characteristics.
For example, we can ask readers to perform small subtasks, such as identifying sources or labeling word choices~\cite{Hamborg2019} to test if these influence their perceptions of bias.

\addition{Another noteworthy aspect here is that our participant group came from Facebook advertising, not from platforms like Amazon Mechanical Turk or Upwork typically used for crowdsourcing. 
This suggests that users outside of crowd work platforms can also perform effectively on crowdsourced tasks.
In the future, research could look into how well workers from crowdsourcing platforms and other sources compare in terms of performance.}
 
Our result indicates that crowd annotation in subjective tasks is to a little extent affected by users' backgrounds---in our case, their news expertise, aligning with prior works~\cite{hube2019understanding,roitero2020can,draws2022effects}.
Therefore, we can train users using their news expertise as a targeting criterion.
Since user performance also varies by task, helping users improve quality on a particular task area might also help.
Furthermore, designers can support users in annotation tasks through various interventions.
For example, since our TF-IDF models identified some generic words that can distinguish false positives, such data could also be used to provide users with feedback or warnings to improve annotation quality.

\addition{We also discovered the effects of comparative annotation on users' overall impressions, leading to differences in perceptions of viewpoints, information attributes (placement, depth, and factuality/opinion), and emotional attributes (empathetic vs. inflammatory language). 
These differences could impact users' attitudes towards an article.
For example, between informational and emotional attributes, understanding which differences impact perceptions of trustworthiness could be one future avenue of work.}
 
To improve users' performance, one option could be through collaboration---learning from each other through social annotation~\cite{Hill2009}.
Indeed, prior research shows that when people see others' annotations, it can persuade them to take certain actions, such as changing ratings when faced with opposing social opinions~\cite{cosley2003seeing}.
Furthermore, research suggests that displaying social information about the annotator, such as their level of expertise, can persuade and build trust~\cite{golbeck2010trust}.
Incorporating social information on other annotators during collaboration may improve learning.
In a collaborative environment, we still need to handle annotator bias, since bias from a small group of users could propagate to a larger pool of users and cause unexpected effects.
\addition{Therefore, examining such collaborative annotations and their impact on user performance is another potential direction of research.}
  

\subsection{RQ2: The Effect of Engaging through Annotation}\label{sec-dis2}
While it may not be true in all cases, our results indicate that in cases where there is significant contrast between a pair of news articles, users might be somewhat influenced by comparative annotation tasks.
Our work can inform related future works on improving engagement with plural viewpoints through annotations~\cite{Wood2018}.
Compared to works that show visualizing biases alone does not improve perception of bias~\cite{spinde2020enabling}, our work suggests that additional engagement could be helpful.
The effect we see may stem from complex information processing that occurs when users engage with competing messages~\cite{bennett1981perception}.
Since our result did not reveal any universally significant effect, it does point towards the idea that only certain perceptions are affected.
Therefore, one direction for future research could include looking into different perception paradigms to further identify the limits of such effects.

\addition{Even though we found limited effects on perceptions of credibility, this does not necessarily limit the applicability of comparative annotation. As we saw in section 5.1.6, a user's understanding of the differences between articles could have other impacts.
Besides, repeatedly annotating two sources can create certain impressions in the long run.
For example, seeing repeated differences in the use of factual statements or depth in reporting could affect users' perceptions of credibility.
Furthermore, we can ask whether crowdworkers from such platforms as Mechanical Turk would also remain unaffected by the annotation task.
In any case, \sys{} could be purposefully deployed to crowdworkers while also providing general users the option to perform annotation.
In such a case, users desiring more ways to engage and community fact-checkers might be more attracted to it.
Regardless, there are further uses for the annotated data.
}

\subsection{Applications of Annotated Data}
Our annotated data could be used in various ways.
For instance, it could be incorporated into a system that combines information from multiple sources to provide a holistic view of an event.
It is not uncommon in online spaces for information overload to make it harder for people to efficiently consume information~\cite{eppler2008concept,aldoory2006roles}.
A holistic view could particularly be useful to users in such a scenario, especially for sensemaking purposes~\cite{weick2005organizing,weick1995sensemaking}.
Such a system would mimic strategies humans typically employ to consume information efficiently, such as organizing information by tagging, sorting, and indexing~\cite{carver2007human}.
We could further build upon this by introducing mechanisms for peer-curated information~\cite{pollar2003surviving}
A second potential use of the annotated data would be in training algorithmic models to generate better annotations, which could in turn be used to better curate information for readers.
As we saw in our initial think-aloud interviews, people find the accuracy of existing SOTA ML systems insufficient for finding semantic similarities and differences.
The annotated data could help to improve such algorithms.
\addition{A third use of the data is for fact-checkers.
Fact-checkers can use annotated information to validate claims through the use of linked statements from multiple sources.
They can also use such links to trace the origins of statements.
Perhaps a portion of these fact-checking tasks could be delegated to automatic fact-checking algorithms.
Furthermore, even crowd fact-checkers (e.g., from Twitter's BirdWatch) could use the annotated data to validate claims.}

 

\subsection{Merging Articles Into One and Testing Effects}
\addition{
One of the goals of this research on comparative annotation was to combine diverse perspectives into one.
With our annotated data, crowd tasks can be designed to accomplish such merging of perspectives.
However, there are some considerations for task design in this process.
Take similar statements as an example---if two statements are very similar, a task could ask workers to choose one or the other.
On the other hand, if selecting one statement necessarily results in the omission of important information from the other, then the crowd task may also require editing.
In the case of merging important disparate statements, as noted in our think-aloud interviews, one important consideration is checking whether a statement fits the narrative of the current article.
We can either include this criteria or discard it, which would lead to differences in the outcome, (i.e, the merged article).
Taking a step further, this merging process can be extended from article pairs to larger groups of articles.
Merging larger groups of articles would require a multistep selection, voting, and reconciliation process.
Finally, while we found that the effect of annotation on perception was limited, could merged articles affect users' perceptions of an event differently than articles from a single source?
Future research answering such a question would generate new knowledge regarding the utility of comparative annotation.
}

\subsection{Implications for Comparative Annotation Task Design}
\addition{Motivated by users' perceptions of \sys, we identified two major issues in the comparative annotation task: the lengthy nature of the task and difficulty in performing the task.
Since one of our research questions focused on the impact of performing annotation, our experiment was designed so that users performed a complete annotation task on two articles before responding to the questions.
If the annotation impact is not of interest, both of these issues can be resolved. 
First, we can modularize the tasks by breaking them into small pieces (e.g., making connections between two paragraphs instead of two entire articles), in line with prior research on devising microtasks for complex work~\cite{kittur2008crowdsourcing,kittur2011crowdforge}.
However, could such modularization cause a backfire effect?
For example, if an annotator is assigned two dissimilar paragraphs from a pair of broadly very similar articles, could that skew their perception?
This is one potential consideration for designing small, modular tasks.}

\addition{Second, even if the task is not divided into smaller components, there are other options for improvement. For instance, finding similarities can be made easier through the addition of such features as automatic suggestion and filtering.
Here, algorithms can provide automatic suggestions and users can search by keyword to limit the options to choose from.}

\addition{Third, tasks can be divided for co-annotation to reduce difficulty.
For example, one annotator might suggest connections while another annotator votes on the suggestions.
In addition, as a tutorial, displaying example annotations from other users could also help resolve some concerns.
However, the examples need to be generic enough not to significantly impact users' own future annotations.}

\addition{Fourth, apart from issues related to task difficulty, there is another issue that will need attention in the future.
In the think-aloud interviews and the deployment of \sys{}, we discovered some disconnects in the rationales provided for annotations.
Particularly, for connection-making, we did not see use of thematic similarity during deployment.
Perhaps regular users may need nudges to identify high-level thematic similarity.
Overall, there are ample opportunities for improving the tasks in \sys.
}


\subsection{Limitations}
Our work is not without limitations. First, our study was conducted in a controlled environment which may differ from that of a user's typical news reading sessions. Therefore, some of the observed effects could have been products of the environment. However, we emulated a typical news consumption environment as best we could, from content selection to the design of the interface. Therefore, our results offer some validity that future works can build on.
Second, since our study procedure involved signing up for the study and voluntary completion criteria, some self-selection bias exists, similar to other research in this domain. However, we did advertise on Facebook to find users organically instead of recruiting users from crowd survey platforms, which provided some benefits to the selection process.
Third, we conducted the study within a US-centric context, limiting its generalizability. Future research could resolve such issues by conducting similar research with a larger country pool.
Finally, the task in the study was a bit lengthy (~20 min) relative to tasks that crowd workers typically perform.
Though the articles in the study were not excessively long (11-16 sentences), this could still have affected task quality.
Future work can further examine how performance varies by task complexity.
Overall, our work has certain merits that require further exploration in the future.

%% file: tex/7conclusion.tex
\section{Conclusion}
In this work, we examined how well users perform on a comparative news annotation task featuring a pair of news articles, and how the annotation task affects users' perceptions of the articles.
Comparing our users' annotations against those of experts, we found that users generally performed very poorly on the annotation task.
However, certain information, such as the number of users who made a given annotation and users' rationales behind annotations, can be used to detect incorrect annotations.
Furthermore, we found some marginal changes in users' credibility perceptions for certain news articles after completing the annotation process.
Our work has implications for designing future comparative annotation systems.

\begin{acks}
\addition{This paper would not be possible without our study participants. We also appreciate the valuable feedback we received from the anonymous reviewers, the members of the EchoLab at Virginia Tech, and the members of the SCALE Lab at the University of Washington, and Natasha Noy from Google Research. Bhuiyan and Mitra were generously supported by National Science Foundation grant \#2128642.}
\end{acks}

%% file: tex/8appendix.tex
\balance
\begin{table*}
    \centering
    \small
     \begin{tabular}{llrclrclrclrc}
& \multicolumn{3}{c}{Imp. Recall (M1)} & \multicolumn{3}{c}{Imp. Precision (M2)} & \multicolumn{3}{c}{Conn. Recall(M3)} & \multicolumn{3}{c}{Conn. Precision(M4)} \\
\cline{2-3} \cline{5-6} \cline{8-9} \cline{11-12}
&  $\boldsymbol{\beta}$ & \textbf{std. err.} & &  $\boldsymbol{\beta}$ & \textbf{std. err.}& &  $\boldsymbol{\beta}$ & \textbf{std. err.}& &  $\boldsymbol{\beta}$ & \textbf{std. err.}\\
\hline
(Intercept) & 0.32*** & 0.08 & & 0.21*** & 0.04 & & 0.34*** & 0.05 & & 0.37*** & 0.05 & \\
CEK[Low] & 0.35** & 0.13 & & 0.05 & 0.07 & & 0.07 & 0.06 & & 0.02 & 0.07 & \\
VML[Low] & 0.18 & 0.12 & & -0.02 & 0.06 & & -0.04 & 0.06 & & 0.02 & 0.07 & \\
\hline
 & \multicolumn{2}{c}{$R^2$=0.12} & & \multicolumn{2}{c}{$R^2$=0.01}& & \multicolumn{2}{c}{$R^2$=0.04} & & \multicolumn{2}{c}{$R^2$=0.19}\\
$N_{obs}$=62 & \multicolumn{11}{r}{*p<0.05, **p<0.01, ***p<0.001}\\
\hline
\end{tabular}    \caption{Linear models of recall and precision for connection-making and importance detection with user characteristics as predictors.}
    \label{tab:ann-user}
\end{table*}

\begin{table*}
\centering
\small
\begin{tabular}{llrllr}
& \multicolumn{2}{c}{$Qual$(M5)} & & \multicolumn{2}{c}{$Cred$(M6)}\\
\cline{2-3} \cline{5-6}
& $\boldsymbol{\beta}$ & \textbf{std. err.} & & $\boldsymbol{\beta}$ & \textbf{std. err.}\\
\hline
(Intercept)	& 0.67***	& 0.05 & & 0.65*** &	0.05\\
Group[Treat.]& 0.01 & 0.07 & & 0.11 & 0.07 \\

Article[Abortion(R)] & 0.04 &	0.07 & &	0.01 &	0.07\\
Article[Immigration(L)] &	0.03 &	0.07 & &	0.11 &	0.07 \\
Article[Immigration(R)] & -0.01 & 0.05 & &	0.04 &	0.05\\
Group [Treat.] * Article[Abortion(R)] &	-0.07 &	0.09 & &	-0.17* &	0.08 \\
Group [Treat.] * [Immigration(L)] &	0.00 &	0.09 &	&	-0.18 &	0.09 \\
Group [Treat.] * Article[Immigration(R)] &	0.02 &	0.08 & &	-0.12 &	0.08 \\
\hline
 & \multicolumn{2}{c}{$R^2=0.43$} & & \multicolumn{2}{c}{$R^2=0.48$}\\
$N_{user}=109$, $N_{article}=4$, $N_{obs}=218$ & \multicolumn{5}{r}{*p<0.05, **p<0.01, ***p<0.001}\\
\hline
\end{tabular}
\caption{Mixed-effects regression on quality and credibility score using the interaction of experimental condition and articles.}\label{tab:rq1}
\end{table*}

\appendix

\section{Thinkaloud Interviews Questionnaires}
\label{app-think}

\begin{itemize}
    \item What are the viewpoints expressed in each article? How would you compare the viewpoints between the articles?
    \item How would you compare the numbers of actors reported in each article?
    \item How would you compare each article providing complete information about what happened/where/when/who was involved?
    \item How would you compare the analytical quality in each article? Do they provide information on causes, consequences, evaluations and claims of/from the event?
    \item How transparent are the authors for each article about their sources (e.g,name, function, circumstances of quote)?
    \item How would you compare the comprehensibility of the article pair (e.g., simplicity in terms/phasing, conciseness, coherence)?
    \item How would you compare impartiality in content presentation between the articles (balanced viewpoints and actors, article \item author personally evaluating/judging the reported situation)?
    \item How would you compare ethical standards (e.g., discriminating any party involved, neutral phrasing) between the reports?
    \item Whose perspective this article represents more than others? Is there any particular group/party/side that the article focus to represent compared to the other article?
\end{itemize}

\section{Articles Used in the Deployment}
\label{articles}

\begin{itemize}
    \item $E_1L$: \href{https://web.archive.org/web/20220629151141/https://www.thedailybeast.com/san-antonio-texas-human-smuggling-tragedy-sees-three-arrests}{\textcolor{blue}{Immigration (Left)}}
    \item $E_1R$: \href{https://web.archive.org/web/20220629225210/https://www.foxnews.com/us/texas-tractor-trailer-driver-smuggling-migrants-arrested-very-high-meth}{\textcolor{blue}{Immigration (Right)}}
    \item $E_2L$: \href{https://web.archive.org/web/20220802191541/https://www.usatoday.com/story/news/politics/2022/08/02/doj-challenges-idaho-abortion-law/10215951002/}{\textcolor{blue}{Abortion (Left)}}
    \item $E_2R$: \href{https://web.archive.org/web/20220907150710/https://www.foxnews.com/politics/idaho-abortion-ban-challenged-justice-department}{\textcolor{blue}{Abortion (Right)}}
\end{itemize}

\section{Effect of User Characteristics}\label{app:user-char}
\addition{We modeled user characteristics to predict precision and recall in annotation tasks, shown in table~\ref{tab:ann-user}.
We accounted for several factors in these models, including users' demographic characteristics (age, gender, education, and political affiliation) and news expertise metrics (CEK and VML).}

\section{RQ2: Mixed-Effects Models}\label{app-rq2}
\addition{Besides ANOVA, we also performed a series of mixed-effects regression model on users' quality and credibility perception using experimental variables, in Table~\ref{tab:rq1}. Similar to ANOVA results, we found significant interaction effect on credibility only for high contrast article.}

%% file: main.bbl

\begin{thebibliography}{84}


\ifx \showCODEN    \undefined \def \showCODEN     #1{\unskip}     \fi
\ifx \showDOI      \undefined \def \showDOI       #1{#1}\fi
\ifx \showISBNx    \undefined \def \showISBNx     #1{\unskip}     \fi
\ifx \showISBNxiii \undefined \def \showISBNxiii  #1{\unskip}     \fi
\ifx \showISSN     \undefined \def \showISSN      #1{\unskip}     \fi
\ifx \showLCCN     \undefined \def \showLCCN      #1{\unskip}     \fi
\ifx \shownote     \undefined \def \shownote      #1{#1}          \fi
\ifx \showarticletitle \undefined \def \showarticletitle #1{#1}   \fi
\ifx \showURL      \undefined \def \showURL       {\relax}        \fi
\providecommand\bibfield[2]{#2}
\providecommand\bibinfo[2]{#2}
\providecommand\natexlab[1]{#1}
\providecommand\showeprint[2][]{arXiv:#2}

\bibitem[Aldoory and Van~Dyke(2006)]%
        {aldoory2006roles}
\bibfield{author}{\bibinfo{person}{Linda Aldoory} {and} \bibinfo{person}{Mark~A
  Van~Dyke}.} \bibinfo{year}{2006}\natexlab{}.
\newblock \showarticletitle{The roles of perceived “shared” involvement and
  information overload in understanding how audiences make meaning of news
  about bioterrorism}.
\newblock \bibinfo{journal}{\emph{Journalism \& Mass Communication Quarterly}}
  \bibinfo{volume}{83}, \bibinfo{number}{2} (\bibinfo{year}{2006}),
  \bibinfo{pages}{346--361}.
\newblock


\bibitem[Allen et~al\mbox{.}(2021)]%
        {allen2021scaling}
\bibfield{author}{\bibinfo{person}{Jennifer Allen}, \bibinfo{person}{Antonio~A
  Arechar}, \bibinfo{person}{Gordon Pennycook}, {and} \bibinfo{person}{David~G
  Rand}.} \bibinfo{year}{2021}\natexlab{}.
\newblock \showarticletitle{Scaling up fact-checking using the wisdom of
  crowds}.
\newblock \bibinfo{journal}{\emph{Science advances}} \bibinfo{volume}{7},
  \bibinfo{number}{36} (\bibinfo{year}{2021}), \bibinfo{pages}{eabf4393}.
\newblock


\bibitem[Allen et~al\mbox{.}(2022)]%
        {allen2022birds}
\bibfield{author}{\bibinfo{person}{Jennifer Allen}, \bibinfo{person}{Cameron
  Martel}, {and} \bibinfo{person}{David~G Rand}.}
  \bibinfo{year}{2022}\natexlab{}.
\newblock \showarticletitle{Birds of a feather don't fact-check each other:
  Partisanship and the evaluation of news in Twitter's Birdwatch crowdsourced
  fact-checking program}. In \bibinfo{booktitle}{\emph{CHI Conference on Human
  Factors in Computing Systems}}. \bibinfo{pages}{1--19}.
\newblock


\bibitem[Arthur(1994)]%
        {arthur1994inductive}
\bibfield{author}{\bibinfo{person}{W~Brian Arthur}.}
  \bibinfo{year}{1994}\natexlab{}.
\newblock \showarticletitle{Inductive reasoning and bounded rationality}.
\newblock \bibinfo{journal}{\emph{The American economic review}}
  \bibinfo{volume}{84}, \bibinfo{number}{2} (\bibinfo{year}{1994}),
  \bibinfo{pages}{406--411}.
\newblock


\bibitem[Babakar(2018)]%
        {Babakar_2018}
\bibfield{author}{\bibinfo{person}{Mevan Babakar}.}
  \bibinfo{year}{2018}\natexlab{}.
\newblock \bibinfo{title}{Crowdsourced Factchecking}.
\newblock
\newblock


\bibitem[Baker and Edmonds(2021)]%
        {baker2021immigration}
\bibfield{author}{\bibinfo{person}{Joseph~O Baker} {and} \bibinfo{person}{Amy~E
  Edmonds}.} \bibinfo{year}{2021}\natexlab{}.
\newblock \showarticletitle{Immigration, presidential politics, and partisan
  polarization among the American public, 1992--2018}.
\newblock \bibinfo{journal}{\emph{Sociological Spectrum}} \bibinfo{volume}{41},
  \bibinfo{number}{4} (\bibinfo{year}{2021}), \bibinfo{pages}{287--303}.
\newblock


\bibitem[Bateman et~al\mbox{.}(2006)]%
        {bateman2006oats}
\bibfield{author}{\bibinfo{person}{Scott Bateman}, \bibinfo{person}{Rosta
  Farzan}, \bibinfo{person}{Peter Brusilovsky}, {and} \bibinfo{person}{Gord
  McCalla}.} \bibinfo{year}{2006}\natexlab{}.
\newblock \showarticletitle{OATS: The open annotation and tagging system}.
\newblock \bibinfo{journal}{\emph{Proceedings of I2LOR}}
  (\bibinfo{year}{2006}).
\newblock


\bibitem[Bennett(1981)]%
        {bennett1981perception}
\bibfield{author}{\bibinfo{person}{W~Lance Bennett}.}
  \bibinfo{year}{1981}\natexlab{}.
\newblock \showarticletitle{Perception and cognition}.
\newblock In \bibinfo{booktitle}{\emph{The handbook of political behavior}}.
  \bibinfo{publisher}{Springer}, \bibinfo{pages}{69--193}.
\newblock


\bibitem[Bhuiyan et~al\mbox{.}(2020)]%
        {bhuiyan2020investigating}
\bibfield{author}{\bibinfo{person}{Md~Momen Bhuiyan}, \bibinfo{person}{Amy~X
  Zhang}, \bibinfo{person}{Connie~Moon Sehat}, {and} \bibinfo{person}{Tanushree
  Mitra}.} \bibinfo{year}{2020}\natexlab{}.
\newblock \showarticletitle{Investigating differences in crowdsourced news
  credibility assessment: Raters, tasks, and expert criteria}.
\newblock \bibinfo{journal}{\emph{Proceedings of the ACM on Human-Computer
  Interaction}} \bibinfo{volume}{4}, \bibinfo{number}{CSCW2}
  (\bibinfo{year}{2020}), \bibinfo{pages}{1--26}.
\newblock


\bibitem[Br{\aa}ten and Str{\o}ms{\o}(2011)]%
        {braaten2011measuring}
\bibfield{author}{\bibinfo{person}{Ivar Br{\aa}ten} {and}
  \bibinfo{person}{Helge~I Str{\o}ms{\o}}.} \bibinfo{year}{2011}\natexlab{}.
\newblock \showarticletitle{Measuring strategic processing when students read
  multiple texts}.
\newblock \bibinfo{journal}{\emph{Metacognition and Learning}}
  \bibinfo{volume}{6}, \bibinfo{number}{2} (\bibinfo{year}{2011}),
  \bibinfo{pages}{111--130}.
\newblock


\bibitem[Budescu and Chen(2015)]%
        {budescu2015identifying}
\bibfield{author}{\bibinfo{person}{David~V Budescu} {and} \bibinfo{person}{Eva
  Chen}.} \bibinfo{year}{2015}\natexlab{}.
\newblock \showarticletitle{Identifying expertise to extract the wisdom of
  crowds}.
\newblock \bibinfo{journal}{\emph{Management Science}} \bibinfo{volume}{61},
  \bibinfo{number}{2} (\bibinfo{year}{2015}), \bibinfo{pages}{267--280}.
\newblock


\bibitem[Carver and Turoff(2007)]%
        {carver2007human}
\bibfield{author}{\bibinfo{person}{Liz Carver} {and} \bibinfo{person}{Murray
  Turoff}.} \bibinfo{year}{2007}\natexlab{}.
\newblock \showarticletitle{Human-computer interaction: the human and computer
  as a team in emergency management information systems}.
\newblock \bibinfo{journal}{\emph{Commun. ACM}} \bibinfo{volume}{50},
  \bibinfo{number}{3} (\bibinfo{year}{2007}), \bibinfo{pages}{33--38}.
\newblock


\bibitem[Center(2014)]%
        {center2014political}
\bibfield{author}{\bibinfo{person}{Pew~Reseach Center}.}
  \bibinfo{year}{2014}\natexlab{}.
\newblock \showarticletitle{Political polarization in the american public}.
\newblock \bibinfo{journal}{\emph{Ann Rev Polit Sci}} (\bibinfo{year}{2014}).
\newblock


\bibitem[Chhabra and Resnick(2013)]%
        {chhabra2013does}
\bibfield{author}{\bibinfo{person}{Sidharth Chhabra} {and}
  \bibinfo{person}{Paul Resnick}.} \bibinfo{year}{2013}\natexlab{}.
\newblock \showarticletitle{Does clustered presentation lead readers to diverse
  selections?}
\newblock In \bibinfo{booktitle}{\emph{CHI'13 Extended Abstracts on Human
  Factors in Computing Systems}}. \bibinfo{pages}{1689--1694}.
\newblock


\bibitem[Chi(1992)]%
        {chi1992conceptual}
\bibfield{author}{\bibinfo{person}{Michelene Chi}.}
  \bibinfo{year}{1992}\natexlab{}.
\newblock \showarticletitle{Conceptual change within and across ontological
  categories: Examples from learning and discovery in science}.
\newblock  (\bibinfo{year}{1992}).
\newblock


\bibitem[Choi(2015)]%
        {choi2015two}
\bibfield{author}{\bibinfo{person}{Sujin Choi}.}
  \bibinfo{year}{2015}\natexlab{}.
\newblock \showarticletitle{The two-step flow of communication in Twitter-based
  public forums}.
\newblock \bibinfo{journal}{\emph{Social science computer review}}
  \bibinfo{volume}{33}, \bibinfo{number}{6} (\bibinfo{year}{2015}),
  \bibinfo{pages}{696--711}.
\newblock


\bibitem[Correa et~al\mbox{.}(2015)]%
        {correa2015many}
\bibfield{author}{\bibinfo{person}{Denzil Correa},
  \bibinfo{person}{Leandro~Ara{\'u}jo Silva}, \bibinfo{person}{Mainack Mondal},
  \bibinfo{person}{Fabr{\'\i}cio Benevenuto}, {and} \bibinfo{person}{Krishna~P
  Gummadi}.} \bibinfo{year}{2015}\natexlab{}.
\newblock \showarticletitle{The many shades of anonymity: Characterizing
  anonymous social media content}. In \bibinfo{booktitle}{\emph{Ninth
  International AAAI Conference on Web and Social Media}}.
\newblock


\bibitem[Cosley et~al\mbox{.}(2003)]%
        {cosley2003seeing}
\bibfield{author}{\bibinfo{person}{Dan Cosley}, \bibinfo{person}{Shyong~K Lam},
  \bibinfo{person}{Istvan Albert}, \bibinfo{person}{Joseph~A Konstan}, {and}
  \bibinfo{person}{John Riedl}.} \bibinfo{year}{2003}\natexlab{}.
\newblock \showarticletitle{Is seeing believing? How recommender system
  interfaces affect users' opinions}. In \bibinfo{booktitle}{\emph{Proceedings
  of the SIGCHI conference on Human factors in computing systems}}.
  \bibinfo{pages}{585--592}.
\newblock


\bibitem[Del~Vicario et~al\mbox{.}(2017)]%
        {del2017modeling}
\bibfield{author}{\bibinfo{person}{Michela Del~Vicario},
  \bibinfo{person}{Antonio Scala}, \bibinfo{person}{Guido Caldarelli},
  \bibinfo{person}{H~Eugene Stanley}, {and} \bibinfo{person}{Walter
  Quattrociocchi}.} \bibinfo{year}{2017}\natexlab{}.
\newblock \showarticletitle{Modeling confirmation bias and polarization}.
\newblock \bibinfo{journal}{\emph{Scientific reports}} \bibinfo{volume}{7},
  \bibinfo{number}{1} (\bibinfo{year}{2017}), \bibinfo{pages}{1--9}.
\newblock


\bibitem[DellaVigna and Kaplan(2007)]%
        {dellavigna2007fox}
\bibfield{author}{\bibinfo{person}{Stefano DellaVigna} {and}
  \bibinfo{person}{Ethan Kaplan}.} \bibinfo{year}{2007}\natexlab{}.
\newblock \showarticletitle{The Fox News effect: Media bias and voting}.
\newblock \bibinfo{journal}{\emph{The Quarterly Journal of Economics}}
  \bibinfo{volume}{122}, \bibinfo{number}{3} (\bibinfo{year}{2007}),
  \bibinfo{pages}{1187--1234}.
\newblock


\bibitem[DeMarzo et~al\mbox{.}(2003)]%
        {demarzo2003persuasion}
\bibfield{author}{\bibinfo{person}{Peter~M DeMarzo}, \bibinfo{person}{Dimitri
  Vayanos}, {and} \bibinfo{person}{Jeffrey Zwiebel}.}
  \bibinfo{year}{2003}\natexlab{}.
\newblock \showarticletitle{Persuasion bias, social influence, and
  unidimensional opinions}.
\newblock \bibinfo{journal}{\emph{The Quarterly journal of economics}}
  \bibinfo{volume}{118}, \bibinfo{number}{3} (\bibinfo{year}{2003}),
  \bibinfo{pages}{909--968}.
\newblock


\bibitem[Draws et~al\mbox{.}(2022)]%
        {draws2022effects}
\bibfield{author}{\bibinfo{person}{Tim Draws}, \bibinfo{person}{David
  La~Barbera}, \bibinfo{person}{Michael Soprano}, \bibinfo{person}{Kevin
  Roitero}, \bibinfo{person}{Davide Ceolin}, \bibinfo{person}{Alessandro
  Checco}, {and} \bibinfo{person}{Stefano Mizzaro}.}
  \bibinfo{year}{2022}\natexlab{}.
\newblock \showarticletitle{The Effects of Crowd Worker Biases in Fact-Checking
  Tasks}. In \bibinfo{booktitle}{\emph{2022 ACM Conference on Fairness,
  Accountability, and Transparency}}. \bibinfo{pages}{2114--2124}.
\newblock


\bibitem[Druckman and Parkin(2005)]%
        {druckman2005impact}
\bibfield{author}{\bibinfo{person}{James~N Druckman} {and}
  \bibinfo{person}{Michael Parkin}.} \bibinfo{year}{2005}\natexlab{}.
\newblock \showarticletitle{The impact of media bias: How editorial slant
  affects voters}.
\newblock \bibinfo{journal}{\emph{The Journal of Politics}}
  \bibinfo{volume}{67}, \bibinfo{number}{4} (\bibinfo{year}{2005}),
  \bibinfo{pages}{1030--1049}.
\newblock


\bibitem[Entman(2007)]%
        {entman2007framing}
\bibfield{author}{\bibinfo{person}{Robert~M Entman}.}
  \bibinfo{year}{2007}\natexlab{}.
\newblock \showarticletitle{Framing bias: Media in the distribution of power}.
\newblock \bibinfo{journal}{\emph{Journal of communication}}
  \bibinfo{volume}{57}, \bibinfo{number}{1} (\bibinfo{year}{2007}),
  \bibinfo{pages}{163--173}.
\newblock


\bibitem[Eppler and Mengis(2008)]%
        {eppler2008concept}
\bibfield{author}{\bibinfo{person}{Martin~J Eppler} {and}
  \bibinfo{person}{Jeanne Mengis}.} \bibinfo{year}{2008}\natexlab{}.
\newblock \showarticletitle{The concept of information overload-a review of
  literature from organization science, accounting, marketing, mis, and related
  disciplines (2004)}.
\newblock \bibinfo{journal}{\emph{Kommunikationsmanagement im Wandel}}
  (\bibinfo{year}{2008}), \bibinfo{pages}{271--305}.
\newblock


\bibitem[Eveland~Jr and Dunwoody(2001)]%
        {eveland2001user}
\bibfield{author}{\bibinfo{person}{William~P Eveland~Jr} {and}
  \bibinfo{person}{Sharon Dunwoody}.} \bibinfo{year}{2001}\natexlab{}.
\newblock \showarticletitle{User control and structural isomorphism or
  disorientation and cognitive load? Learning from the Web versus print}.
\newblock \bibinfo{journal}{\emph{Communication research}}
  \bibinfo{volume}{28}, \bibinfo{number}{1} (\bibinfo{year}{2001}),
  \bibinfo{pages}{48--78}.
\newblock


\bibitem[Fiorina et~al\mbox{.}(2008)]%
        {fiorina2008political}
\bibfield{author}{\bibinfo{person}{Morris~P Fiorina}, \bibinfo{person}{Samuel~J
  Abrams}, {et~al\mbox{.}}} \bibinfo{year}{2008}\natexlab{}.
\newblock \showarticletitle{Political polarization in the American public}.
\newblock \bibinfo{journal}{\emph{ANNUAL REVIEW OF POLITICAL SCIENCE-PALO
  ALTO-}}  \bibinfo{volume}{11} (\bibinfo{year}{2008}), \bibinfo{pages}{563}.
\newblock


\bibitem[Golbeck and Fleischmann(2010)]%
        {golbeck2010trust}
\bibfield{author}{\bibinfo{person}{Jennifer Golbeck} {and}
  \bibinfo{person}{Kenneth~R Fleischmann}.} \bibinfo{year}{2010}\natexlab{}.
\newblock \showarticletitle{Trust in social Q\&A: the impact of text and photo
  cues of expertise}.
\newblock \bibinfo{journal}{\emph{Proceedings of the American Society for
  Information Science and Technology}} \bibinfo{volume}{47},
  \bibinfo{number}{1} (\bibinfo{year}{2010}), \bibinfo{pages}{1--10}.
\newblock


\bibitem[Hamborg et~al\mbox{.}(2019)]%
        {Hamborg2019}
\bibfield{author}{\bibinfo{person}{Felix Hamborg}, \bibinfo{person}{Karsten
  Donnay}, {and} \bibinfo{person}{Bela Gipp}.} \bibinfo{year}{2019}\natexlab{}.
\newblock \showarticletitle{{Automated identification of media bias in news
  articles: an interdisciplinary literature review}}.
\newblock \bibinfo{journal}{\emph{International Journal on Digital Libraries}}
  \bibinfo{volume}{20}, \bibinfo{number}{4} (\bibinfo{year}{2019}),
  \bibinfo{pages}{391--415}.
\newblock
\showISSN{14321300}


\bibitem[Hassan et~al\mbox{.}(2017)]%
        {Hassan:2017:TAS}
\bibfield{author}{\bibinfo{person}{Naeemul Hassan}, \bibinfo{person}{Mohammad
  Yousuf}, \bibinfo{person}{Mahfuzul Haque}, \bibinfo{person}{Javier
  A~Suarez~Rivas}, {and} \bibinfo{person}{Md~Khadimul Islam}.}
  \bibinfo{year}{2017}\natexlab{}.
\newblock \showarticletitle{Towards A Sustainable Model for Fact-checking
  Platforms: Examining the Roles of Automation, Crowds and Professionals}.
\newblock
\urldef\tempurl%
\url{https://doi.org/10.1145/3308560.3316734}
\showDOI{\tempurl}


\bibitem[Herman and Chomsky(2010)]%
        {herman2010manufacturing}
\bibfield{author}{\bibinfo{person}{Edward~S Herman} {and} \bibinfo{person}{Noam
  Chomsky}.} \bibinfo{year}{2010}\natexlab{}.
\newblock \bibinfo{booktitle}{\emph{Manufacturing consent: The political
  economy of the mass media}}.
\newblock \bibinfo{publisher}{Random House}.
\newblock


\bibitem[Hill et~al\mbox{.}(2009)]%
        {Hill2009}
\bibfield{author}{\bibinfo{person}{Janette~R. Hill}, \bibinfo{person}{Liyan
  Song}, {and} \bibinfo{person}{Richard~E. West}.}
  \bibinfo{year}{2009}\natexlab{}.
\newblock \showarticletitle{{Social learning theory and web-based learning
  environments: A review of research and discussion of implications}}.
\newblock \bibinfo{journal}{\emph{International Journal of Phytoremediation}}
  \bibinfo{volume}{21}, \bibinfo{number}{1} (\bibinfo{year}{2009}),
  \bibinfo{pages}{88--103}.
\newblock
\showISSN{15497879}


\bibitem[Hube et~al\mbox{.}(2019)]%
        {hube2019understanding}
\bibfield{author}{\bibinfo{person}{Christoph Hube}, \bibinfo{person}{Besnik
  Fetahu}, {and} \bibinfo{person}{Ujwal Gadiraju}.}
  \bibinfo{year}{2019}\natexlab{}.
\newblock \showarticletitle{Understanding and mitigating worker biases in the
  crowdsourced collection of subjective judgments}. In
  \bibinfo{booktitle}{\emph{Proceedings of the 2019 CHI Conference on Human
  Factors in Computing Systems}}. \bibinfo{pages}{1--12}.
\newblock


\bibitem[Institute(2022)]%
        {Overview24online}
\bibfield{author}{\bibinfo{person}{Reuters Institute}.}
  \bibinfo{year}{2022}\natexlab{}.
\newblock \bibinfo{title}{Overview and key findings of the 2022 Digital News
  Report | Reuters Institute for the Study of Journalism}.
\newblock
  \bibinfo{howpublished}{\url{https://reutersinstitute.politics.ox.ac.uk/digital-news-report/2022/dnr-executive-summary}}.
\newblock
\newblock
\shownote{(Accessed on 09/11/2022)}.


\bibitem[Jamieson and Cappella(2008)]%
        {jamieson2008echo}
\bibfield{author}{\bibinfo{person}{Kathleen~Hall Jamieson} {and}
  \bibinfo{person}{Joseph~N Cappella}.} \bibinfo{year}{2008}\natexlab{}.
\newblock \bibinfo{booktitle}{\emph{Echo chamber: Rush Limbaugh and the
  conservative media establishment}}.
\newblock \bibinfo{publisher}{Oxford University Press}.
\newblock


\bibitem[Jurkowitz et~al\mbox{.}(2020)]%
        {USMediaP92online}
\bibfield{author}{\bibinfo{person}{Mark Jurkowitz}, \bibinfo{person}{Amy
  Mitchell}, \bibinfo{person}{Elisa Shearer}, {and} \bibinfo{person}{Mason
  Walker}.} \bibinfo{year}{2020}\natexlab{}.
\newblock \bibinfo{title}{U.S. Media Polarization and the 2020 Election: A
  Nation Divided | Pew Research Center}.
\newblock
  \bibinfo{howpublished}{\url{https://www.pewresearch.org/journalism/2020/01/24/u-s-media-polarization-and-the-2020-election-a-nation-divided/}}.
\newblock
\newblock
\shownote{(Accessed on 09/15/2022)}.


\bibitem[Kahneman and Tversky(2013)]%
        {kahneman2013choices}
\bibfield{author}{\bibinfo{person}{Daniel Kahneman} {and} \bibinfo{person}{Amos
  Tversky}.} \bibinfo{year}{2013}\natexlab{}.
\newblock \showarticletitle{Choices, values, and frames}.
\newblock In \bibinfo{booktitle}{\emph{Handbook of the fundamentals of
  financial decision making: Part I}}. \bibinfo{publisher}{World Scientific},
  \bibinfo{pages}{269--278}.
\newblock


\bibitem[Kavale(1980)]%
        {kavale1980reasoning}
\bibfield{author}{\bibinfo{person}{Kenneth~A Kavale}.}
  \bibinfo{year}{1980}\natexlab{}.
\newblock \showarticletitle{The reasoning abilities of normal and learning
  disabled readers on measures of reading comprehension}.
\newblock \bibinfo{journal}{\emph{Learning Disability Quarterly}}
  \bibinfo{volume}{3}, \bibinfo{number}{4} (\bibinfo{year}{1980}),
  \bibinfo{pages}{34--45}.
\newblock


\bibitem[Kawase et~al\mbox{.}(2009)]%
        {kawase2009comparison}
\bibfield{author}{\bibinfo{person}{Ricardo Kawase}, \bibinfo{person}{Eelco
  Herder}, {and} \bibinfo{person}{Wolfgang Nejdl}.}
  \bibinfo{year}{2009}\natexlab{}.
\newblock \showarticletitle{A comparison of paper-based and online annotations
  in the workplace}. In \bibinfo{booktitle}{\emph{European Conference on
  Technology Enhanced Learning}}. Springer, \bibinfo{pages}{240--253}.
\newblock


\bibitem[Kiesow et~al\mbox{.}(2021)]%
        {Kiesow2021}
\bibfield{author}{\bibinfo{person}{Damon Kiesow}, \bibinfo{person}{Shuhua
  Zhou}, {and} \bibinfo{person}{Lei Guo}.} \bibinfo{year}{2021}\natexlab{}.
\newblock \showarticletitle{{Affordances for Sense-Making: Exploring Their
  Availability for Users of Online News Sites}}.
\newblock \bibinfo{journal}{\emph{Digital Journalism}} \bibinfo{volume}{0},
  \bibinfo{number}{0} (\bibinfo{year}{2021}), \bibinfo{pages}{1--20}.
\newblock
\showISSN{2167082X}
\urldef\tempurl%
\url{https://doi.org/10.1080/21670811.2021.1989316}
\showDOI{\tempurl}


\bibitem[Kittur et~al\mbox{.}(2008)]%
        {kittur2008crowdsourcing}
\bibfield{author}{\bibinfo{person}{Aniket Kittur}, \bibinfo{person}{Ed~H Chi},
  {and} \bibinfo{person}{Bongwon Suh}.} \bibinfo{year}{2008}\natexlab{}.
\newblock \showarticletitle{Crowdsourcing user studies with Mechanical Turk}.
  In \bibinfo{booktitle}{\emph{Proceedings of the SIGCHI conference on human
  factors in computing systems}}. \bibinfo{pages}{453--456}.
\newblock


\bibitem[Kittur et~al\mbox{.}(2011)]%
        {kittur2011crowdforge}
\bibfield{author}{\bibinfo{person}{Aniket Kittur}, \bibinfo{person}{Boris
  Smus}, \bibinfo{person}{Susheel Khamkar}, {and} \bibinfo{person}{Robert~E
  Kraut}.} \bibinfo{year}{2011}\natexlab{}.
\newblock \showarticletitle{Crowdforge: Crowdsourcing complex work}. In
  \bibinfo{booktitle}{\emph{Proceedings of the 24th annual ACM symposium on
  User interface software and technology}}. \bibinfo{pages}{43--52}.
\newblock


\bibitem[Klayman(1995)]%
        {klayman1995varieties}
\bibfield{author}{\bibinfo{person}{Joshua Klayman}.}
  \bibinfo{year}{1995}\natexlab{}.
\newblock \showarticletitle{Varieties of confirmation bias}.
\newblock \bibinfo{journal}{\emph{Psychology of learning and motivation}}
  \bibinfo{volume}{32} (\bibinfo{year}{1995}), \bibinfo{pages}{385--418}.
\newblock


\bibitem[Kull et~al\mbox{.}(2003)]%
        {kull2003misperceptions}
\bibfield{author}{\bibinfo{person}{Steven Kull}, \bibinfo{person}{Clay Ramsay},
  {and} \bibinfo{person}{Evan Lewis}.} \bibinfo{year}{2003}\natexlab{}.
\newblock \showarticletitle{Misperceptions, the media, and the Iraq war}.
\newblock \bibinfo{journal}{\emph{Political science quarterly}}
  \bibinfo{volume}{118}, \bibinfo{number}{4} (\bibinfo{year}{2003}),
  \bibinfo{pages}{569--598}.
\newblock


\bibitem[Lazarsfeld et~al\mbox{.}(1968)]%
        {lazarsfeld1968people}
\bibfield{author}{\bibinfo{person}{Paul~F Lazarsfeld}, \bibinfo{person}{Bernard
  Berelson}, {and} \bibinfo{person}{Hazel Gaudet}.}
  \bibinfo{year}{1968}\natexlab{}.
\newblock \showarticletitle{The people's choice}.
\newblock In \bibinfo{booktitle}{\emph{The people's choice}}.
  \bibinfo{publisher}{Columbia University Press}.
\newblock


\bibitem[Lebow and Lick(2005)]%
        {lebow2005hylighter}
\bibfield{author}{\bibinfo{person}{David~G Lebow} {and} \bibinfo{person}{Dale~W
  Lick}.} \bibinfo{year}{2005}\natexlab{}.
\newblock \showarticletitle{HyLighter: An effective interactive annotation
  innovation for distance education}. In \bibinfo{booktitle}{\emph{20th Annual
  Conference on Distance Teaching and Learning}}. \bibinfo{pages}{1--5}.
\newblock


\bibitem[Maksl et~al\mbox{.}(2015)]%
        {maksl2015measuring}
\bibfield{author}{\bibinfo{person}{Adam Maksl}, \bibinfo{person}{Seth Ashley},
  {and} \bibinfo{person}{Stephanie Craft}.} \bibinfo{year}{2015}\natexlab{}.
\newblock \showarticletitle{Measuring news media literacy}.
\newblock \bibinfo{journal}{\emph{Journal of Media Literacy Education}}
  \bibinfo{volume}{6}, \bibinfo{number}{3} (\bibinfo{year}{2015}),
  \bibinfo{pages}{29--45}.
\newblock


\bibitem[Metzger et~al\mbox{.}(2015)]%
        {metzger2015cognitive}
\bibfield{author}{\bibinfo{person}{Miriam~J Metzger}, \bibinfo{person}{Ethan~H
  Hartsell}, {and} \bibinfo{person}{Andrew~J Flanagin}.}
  \bibinfo{year}{2015}\natexlab{}.
\newblock \showarticletitle{Cognitive dissonance or credibility? A comparison
  of two theoretical explanations for selective exposure to partisan news}.
\newblock \bibinfo{journal}{\emph{Communication Research}}
  (\bibinfo{year}{2015}), \bibinfo{pages}{0093650215613136}.
\newblock


\bibitem[Meyer(1988)]%
        {meyer1988defining}
\bibfield{author}{\bibinfo{person}{Philip Meyer}.}
  \bibinfo{year}{1988}\natexlab{}.
\newblock \showarticletitle{Defining and measuring credibility of newspapers:
  Developing an index}.
\newblock \bibinfo{journal}{\emph{Journalism quarterly}} \bibinfo{volume}{65},
  \bibinfo{number}{3} (\bibinfo{year}{1988}), \bibinfo{pages}{567--574}.
\newblock


\bibitem[Mitra and Gilbert(2015)]%
        {Mitra_Gilbert_2015}
\bibfield{author}{\bibinfo{person}{Tanushree Mitra} {and} \bibinfo{person}{Eric
  Gilbert}.} \bibinfo{year}{2015}\natexlab{}.
\newblock \showarticletitle{CREDBANK: A Large-Scale Social Media Corpus with
  Associated Credibility Annotations}. In \bibinfo{booktitle}{\emph{Proc.
  ICWSM'15}}.
\newblock


\bibitem[Mullainathan and Shleifer(2002)]%
        {mullainathan2002media}
\bibfield{author}{\bibinfo{person}{Sendhil Mullainathan} {and}
  \bibinfo{person}{Andrei Shleifer}.} \bibinfo{year}{2002}\natexlab{}.
\newblock \bibinfo{title}{Media bias}.
\newblock
\newblock


\bibitem[Munson et~al\mbox{.}(2013)]%
        {munson2013encouraging}
\bibfield{author}{\bibinfo{person}{Sean Munson}, \bibinfo{person}{Stephanie
  Lee}, {and} \bibinfo{person}{Paul Resnick}.} \bibinfo{year}{2013}\natexlab{}.
\newblock \showarticletitle{Encouraging reading of diverse political viewpoints
  with a browser widget}. In \bibinfo{booktitle}{\emph{Proceedings of The
  International AAAI Conference on Web and Social Media}},
  Vol.~\bibinfo{volume}{7}. \bibinfo{pages}{419--428}.
\newblock


\bibitem[Munson and Resnick(2010)]%
        {munson2010presenting}
\bibfield{author}{\bibinfo{person}{Sean~A Munson} {and} \bibinfo{person}{Paul
  Resnick}.} \bibinfo{year}{2010}\natexlab{}.
\newblock \showarticletitle{Presenting diverse political opinions: how and how
  much}. In \bibinfo{booktitle}{\emph{Proceedings of the SIGCHI conference on
  human factors in computing systems}}. \bibinfo{pages}{1457--1466}.
\newblock


\bibitem[Napolitan(1972)]%
        {napolitan1972election}
\bibfield{author}{\bibinfo{person}{Joseph Napolitan}.}
  \bibinfo{year}{1972}\natexlab{}.
\newblock \bibinfo{booktitle}{\emph{The election game and how to win it}}.
\newblock \bibinfo{publisher}{Doubleday}.
\newblock


\bibitem[Nickerson(1998)]%
        {nickerson1998confirmation}
\bibfield{author}{\bibinfo{person}{Raymond~S Nickerson}.}
  \bibinfo{year}{1998}\natexlab{}.
\newblock \showarticletitle{Confirmation bias: A ubiquitous phenomenon in many
  guises}.
\newblock \bibinfo{journal}{\emph{Review of general psychology}}
  \bibinfo{volume}{2}, \bibinfo{number}{2} (\bibinfo{year}{1998}),
  \bibinfo{pages}{175--220}.
\newblock


\bibitem[Novak et~al\mbox{.}(2012)]%
        {Novak2012}
\bibfield{author}{\bibinfo{person}{Elena Novak}, \bibinfo{person}{Rim Razzouk},
  {and} \bibinfo{person}{Tristan~E. Johnson}.} \bibinfo{year}{2012}\natexlab{}.
\newblock \showarticletitle{{The educational use of social annotation tools in
  higher education: A literature review}}.
\newblock \bibinfo{journal}{\emph{Internet and Higher Education}}
  \bibinfo{volume}{15}, \bibinfo{number}{1} (\bibinfo{year}{2012}),
  \bibinfo{pages}{39--49}.
\newblock
\showISSN{10967516}
\urldef\tempurl%
\url{https://doi.org/10.1016/j.iheduc.2011.09.002}
\showDOI{\tempurl}


\bibitem[Pariser(2011)]%
        {pariser2011filter}
\bibfield{author}{\bibinfo{person}{Eli Pariser}.}
  \bibinfo{year}{2011}\natexlab{}.
\newblock \bibinfo{booktitle}{\emph{The filter bubble: How the new personalized
  web is changing what we read and how we think}}.
\newblock \bibinfo{publisher}{Penguin}.
\newblock


\bibitem[Park et~al\mbox{.}(2009)]%
        {park2009newscube}
\bibfield{author}{\bibinfo{person}{Souneil Park}, \bibinfo{person}{Seungwoo
  Kang}, \bibinfo{person}{Sangyoung Chung}, {and} \bibinfo{person}{Junehwa
  Song}.} \bibinfo{year}{2009}\natexlab{}.
\newblock \showarticletitle{{NewsCube: delivering multiple aspects of news to
  mitigate media bias}}. In \bibinfo{booktitle}{\emph{Proceedings of the SIGCHI
  Conference on Human Factors in Computing Systems}}. ACM,
  \bibinfo{pages}{443--452}.
\newblock


\bibitem[Park et~al\mbox{.}(2012)]%
        {Park2012}
\bibfield{author}{\bibinfo{person}{Souneil Park}, \bibinfo{person}{Seungwoo
  Kang}, \bibinfo{person}{Sangyoung Chung}, {and} \bibinfo{person}{Junehwa
  Song}.} \bibinfo{year}{2012}\natexlab{}.
\newblock \showarticletitle{{A computational framework for media bias
  mitigation}}.
\newblock \bibinfo{journal}{\emph{ACM Transactions on Interactive Intelligent
  Systems}} \bibinfo{volume}{2}, \bibinfo{number}{2} (\bibinfo{year}{2012}).
\newblock
\showISSN{21606463}
\urldef\tempurl%
\url{https://doi.org/10.1145/2209310.2209311}
\showDOI{\tempurl}


\bibitem[Park et~al\mbox{.}(2011a)]%
        {Park2011a}
\bibfield{author}{\bibinfo{person}{Souneil Park}, \bibinfo{person}{Minsam Ko},
  \bibinfo{person}{Jungwoo Kim}, \bibinfo{person}{Ho-jin Choi}, {and}
  \bibinfo{person}{Junehwa Song}.} \bibinfo{year}{2011}\natexlab{a}.
\newblock \showarticletitle{{NewsCube2 . 0 : An Exploratory Design of a Social
  News Website for Media Bias Mitigation}}.
\newblock \bibinfo{journal}{\emph{Workshop on Social Recommender Systems}}
  (\bibinfo{year}{2011}), \bibinfo{pages}{1--5}.
\newblock
\urldef\tempurl%
\url{https://pdfs.semanticscholar.org/b87b/f0986b2e9fe34a22ed0c19cfd32ed06857d0.pdf}
\showURL{%
\tempurl}


\bibitem[Park et~al\mbox{.}(2011b)]%
        {Park2011}
\bibfield{author}{\bibinfo{person}{Souneil Park}, \bibinfo{person}{Kyung~Soon
  Lee}, {and} \bibinfo{person}{Junehwa Song}.}
  \bibinfo{year}{2011}\natexlab{b}.
\newblock \showarticletitle{{Contrasting opposing views of news articles on
  contentious issues}}.
\newblock \bibinfo{journal}{\emph{ACL-HLT 2011 - Proceedings of the 49th Annual
  Meeting of the Association for Computational Linguistics: Human Language
  Technologies}}  \bibinfo{volume}{1} (\bibinfo{year}{2011}),
  \bibinfo{pages}{340--349}.
\newblock
\showISBNx{9781932432879}


\bibitem[Pipps et~al\mbox{.}(2009)]%
        {pipps2009information}
\bibfield{author}{\bibinfo{person}{Val Pipps}, \bibinfo{person}{Heather
  Walter}, \bibinfo{person}{Kathleen Endres}, {and} \bibinfo{person}{Patrick
  Tabatcher}.} \bibinfo{year}{2009}\natexlab{}.
\newblock \showarticletitle{Information recall of Internet news: Does design
  make a difference? A pilot study}.
\newblock \bibinfo{journal}{\emph{Journal of Magazine Media}}
  \bibinfo{volume}{11}, \bibinfo{number}{1} (\bibinfo{year}{2009}),
  \bibinfo{pages}{1--20}.
\newblock


\bibitem[Pollar(2003)]%
        {pollar2003surviving}
\bibfield{author}{\bibinfo{person}{Odette Pollar}.}
  \bibinfo{year}{2003}\natexlab{}.
\newblock \bibinfo{booktitle}{\emph{Surviving information overload: how to
  find, filter, and focus on what's important}}.
\newblock \bibinfo{publisher}{Thomson Crisp Learning}.
\newblock


\bibitem[Roitero et~al\mbox{.}(2020)]%
        {roitero2020can}
\bibfield{author}{\bibinfo{person}{Kevin Roitero}, \bibinfo{person}{Michael
  Soprano}, \bibinfo{person}{Shaoyang Fan}, \bibinfo{person}{Damiano Spina},
  \bibinfo{person}{Stefano Mizzaro}, {and} \bibinfo{person}{Gianluca
  Demartini}.} \bibinfo{year}{2020}\natexlab{}.
\newblock \showarticletitle{Can The Crowd Identify Misinformation Objectively?
  The Effects of Judgment Scale and Assessor's Background}. In
  \bibinfo{booktitle}{\emph{Proceedings of the 43rd International ACM SIGIR
  Conference on Research and Development in Information Retrieval}}.
  \bibinfo{pages}{439--448}.
\newblock


\bibitem[Scheufele(2000)]%
        {scheufele2000agenda}
\bibfield{author}{\bibinfo{person}{Dietram~A Scheufele}.}
  \bibinfo{year}{2000}\natexlab{}.
\newblock \showarticletitle{Agenda-setting, priming, and framing revisited:
  Another look at cognitive effects of political communication}.
\newblock \bibinfo{journal}{\emph{Mass communication \& society}}
  \bibinfo{volume}{3}, \bibinfo{number}{2-3} (\bibinfo{year}{2000}),
  \bibinfo{pages}{297--316}.
\newblock


\bibitem[Slate(2013)]%
        {Howpeopl90online}
\bibfield{author}{\bibinfo{person}{Slate}.} \bibinfo{year}{2013}\natexlab{}.
\newblock \bibinfo{title}{How people read online: Why you won't finish this
  article.}
\newblock
  \bibinfo{howpublished}{\url{https://slate.com/technology/2013/06/how-people-read-online-why-you-wont-finish-this-article.html}}.
\newblock
\newblock
\shownote{(Accessed on 09/11/2022)}.


\bibitem[Soffer(2021)]%
        {soffer2021algorithmic}
\bibfield{author}{\bibinfo{person}{Oren Soffer}.}
  \bibinfo{year}{2021}\natexlab{}.
\newblock \showarticletitle{Algorithmic personalization and the two-step flow
  of communication}.
\newblock \bibinfo{journal}{\emph{Communication Theory}} \bibinfo{volume}{31},
  \bibinfo{number}{3} (\bibinfo{year}{2021}), \bibinfo{pages}{297--315}.
\newblock


\bibitem[Spinde et~al\mbox{.}(2020)]%
        {spinde2020enabling}
\bibfield{author}{\bibinfo{person}{Timo Spinde}, \bibinfo{person}{Felix
  Hamborg}, \bibinfo{person}{Karsten Donnay}, \bibinfo{person}{Angelica
  Becerra}, {and} \bibinfo{person}{Bela Gipp}.}
  \bibinfo{year}{2020}\natexlab{}.
\newblock \showarticletitle{Enabling news consumers to view and understand
  biased news coverage: a study on the perception and visualization of media
  bias}. In \bibinfo{booktitle}{\emph{Proceedings of the ACM/IEEE joint
  conference on digital libraries in 2020}}. \bibinfo{pages}{389--392}.
\newblock


\bibitem[Stahl et~al\mbox{.}(1996)]%
        {stahl1996happens}
\bibfield{author}{\bibinfo{person}{Steven~A Stahl}, \bibinfo{person}{Cynthia~R
  Hynd}, \bibinfo{person}{Bruce~K Britton}, \bibinfo{person}{Mary~M McNish},
  {and} \bibinfo{person}{Dennis Bosquet}.} \bibinfo{year}{1996}\natexlab{}.
\newblock \showarticletitle{What happens when students read multiple source
  documents in history?}
\newblock \bibinfo{journal}{\emph{Reading Research Quarterly}}
  \bibinfo{volume}{31}, \bibinfo{number}{4} (\bibinfo{year}{1996}),
  \bibinfo{pages}{430--456}.
\newblock


\bibitem[Stroud(2010)]%
        {stroud2010polarization}
\bibfield{author}{\bibinfo{person}{Natalie~Jomini Stroud}.}
  \bibinfo{year}{2010}\natexlab{}.
\newblock \showarticletitle{Polarization and partisan selective exposure}.
\newblock \bibinfo{journal}{\emph{Journal of communication}}
  \bibinfo{volume}{60}, \bibinfo{number}{3} (\bibinfo{year}{2010}),
  \bibinfo{pages}{556--576}.
\newblock


\bibitem[Stroud(2011)]%
        {stroud2011niche}
\bibfield{author}{\bibinfo{person}{Natalie~Jomini Stroud}.}
  \bibinfo{year}{2011}\natexlab{}.
\newblock \bibinfo{booktitle}{\emph{Niche news: The politics of news choice}}.
\newblock \bibinfo{publisher}{Oxford University Press on Demand}.
\newblock


\bibitem[Sunstein(1999)]%
        {sunstein1999law}
\bibfield{author}{\bibinfo{person}{Cass~R Sunstein}.}
  \bibinfo{year}{1999}\natexlab{}.
\newblock \showarticletitle{The law of group polarization}.
\newblock \bibinfo{journal}{\emph{University of Chicago Law School, John M.
  Olin Law \& Economics Working Paper}} \bibinfo{number}{91}
  (\bibinfo{year}{1999}).
\newblock


\bibitem[Sunstein(2009)]%
        {sunstein2009http}
\bibfield{author}{\bibinfo{person}{Cass~R Sunstein}.}
  \bibinfo{year}{2009}\natexlab{}.
\newblock \bibinfo{title}{http://Republic.com 2.0}.
\newblock
\newblock


\bibitem[Surowiecki(2005)]%
        {surowiecki2005wisdom}
\bibfield{author}{\bibinfo{person}{James Surowiecki}.}
  \bibinfo{year}{2005}\natexlab{}.
\newblock \bibinfo{booktitle}{\emph{The wisdom of crowds}}.
\newblock \bibinfo{publisher}{Anchor}.
\newblock


\bibitem[Tewksbury and Althaus(2000)]%
        {tewksbury2000differences}
\bibfield{author}{\bibinfo{person}{David Tewksbury} {and}
  \bibinfo{person}{Scott~L Althaus}.} \bibinfo{year}{2000}\natexlab{}.
\newblock \showarticletitle{Differences in knowledge acquisition among readers
  of the paper and online versions of a national newspaper}.
\newblock \bibinfo{journal}{\emph{Journalism \& Mass Communication Quarterly}}
  \bibinfo{volume}{77}, \bibinfo{number}{3} (\bibinfo{year}{2000}),
  \bibinfo{pages}{457--479}.
\newblock


\bibitem[Urban and Schweiger(2014)]%
        {urban2014news}
\bibfield{author}{\bibinfo{person}{Juliane Urban} {and}
  \bibinfo{person}{Wolfgang Schweiger}.} \bibinfo{year}{2014}\natexlab{}.
\newblock \showarticletitle{News quality from the recipients' perspective:
  Investigating recipients' ability to judge the normative quality of news}.
\newblock \bibinfo{journal}{\emph{Journalism Studies}} \bibinfo{volume}{15},
  \bibinfo{number}{6} (\bibinfo{year}{2014}), \bibinfo{pages}{821--840}.
\newblock


\bibitem[Vosniadou and Brewer(1987)]%
        {vosniadou1987theories}
\bibfield{author}{\bibinfo{person}{Stella Vosniadou} {and}
  \bibinfo{person}{William~F Brewer}.} \bibinfo{year}{1987}\natexlab{}.
\newblock \showarticletitle{Theories of knowledge restructuring in
  development}.
\newblock \bibinfo{journal}{\emph{Review of educational research}}
  \bibinfo{volume}{57}, \bibinfo{number}{1} (\bibinfo{year}{1987}),
  \bibinfo{pages}{51--67}.
\newblock


\bibitem[Vosniadou and Ortony(1989)]%
        {vosniadou1989similarity}
\bibfield{author}{\bibinfo{person}{Stella Vosniadou} {and}
  \bibinfo{person}{Andrew Ortony}.} \bibinfo{year}{1989}\natexlab{}.
\newblock \bibinfo{booktitle}{\emph{Similarity and analogical reasoning}}.
\newblock \bibinfo{publisher}{Cambridge University Press}.
\newblock


\bibitem[Vraga et~al\mbox{.}(2015)]%
        {vraga2015multi}
\bibfield{author}{\bibinfo{person}{Emily Vraga}, \bibinfo{person}{Melissa
  Tully}, \bibinfo{person}{John~E Kotcher}, \bibinfo{person}{Anne-Bennett
  Smithson}, {and} \bibinfo{person}{Melissa Broeckelman-Post}.}
  \bibinfo{year}{2015}\natexlab{}.
\newblock \showarticletitle{A Multi-Dimensional Approach to Measuring News
  Media Literacy.}
\newblock \bibinfo{journal}{\emph{Journal of Media Literacy Education}}
  \bibinfo{volume}{7}, \bibinfo{number}{3} (\bibinfo{year}{2015}),
  \bibinfo{pages}{41--53}.
\newblock


\bibitem[Weick(1995)]%
        {weick1995sensemaking}
\bibfield{author}{\bibinfo{person}{Karl~E Weick}.}
  \bibinfo{year}{1995}\natexlab{}.
\newblock \bibinfo{booktitle}{\emph{Sensemaking in organizations}}.
  Vol.~\bibinfo{volume}{3}.
\newblock \bibinfo{publisher}{Sage}.
\newblock


\bibitem[Weick et~al\mbox{.}(2005)]%
        {weick2005organizing}
\bibfield{author}{\bibinfo{person}{Karl~E Weick}, \bibinfo{person}{Kathleen~M
  Sutcliffe}, {and} \bibinfo{person}{David Obstfeld}.}
  \bibinfo{year}{2005}\natexlab{}.
\newblock \showarticletitle{Organizing and the process of sensemaking}.
\newblock \bibinfo{journal}{\emph{Organization science}} \bibinfo{volume}{16},
  \bibinfo{number}{4} (\bibinfo{year}{2005}), \bibinfo{pages}{409--421}.
\newblock


\bibitem[Wood et~al\mbox{.}(2018)]%
        {Wood2018}
\bibfield{author}{\bibinfo{person}{Gavin Wood}, \bibinfo{person}{Kiel Long},
  \bibinfo{person}{Tom Feltwell}, \bibinfo{person}{Scarlett Rowland},
  \bibinfo{person}{Phillip Brooker}, \bibinfo{person}{Jamie Mahoney},
  \bibinfo{person}{John Vines}, \bibinfo{person}{Julie Barnett}, {and}
  \bibinfo{person}{Shaun Lawson}.} \bibinfo{year}{2018}\natexlab{}.
\newblock \showarticletitle{{Rethinking engagement with online news through
  social and visual co-annotation}}.
\newblock \bibinfo{journal}{\emph{Conference on Human Factors in Computing
  Systems - Proceedings}}  \bibinfo{volume}{2018-April} (\bibinfo{year}{2018}),
  \bibinfo{pages}{1--12}.
\newblock
\showISBNx{9781450356206}
\urldef\tempurl%
\url{https://doi.org/10.1145/3173574.3174150}
\showDOI{\tempurl}


\bibitem[Zaller et~al\mbox{.}(1992)]%
        {zaller1992nature}
\bibfield{author}{\bibinfo{person}{John~R Zaller} {et~al\mbox{.}}}
  \bibinfo{year}{1992}\natexlab{}.
\newblock \bibinfo{booktitle}{\emph{The nature and origins of mass opinion}}.
\newblock \bibinfo{publisher}{Cambridge university press}.
\newblock


\bibitem[Zhang and Soergel(2020)]%
        {Zhang2020}
\bibfield{author}{\bibinfo{person}{Pengyi Zhang} {and}
  \bibinfo{person}{Dagobert Soergel}.} \bibinfo{year}{2020}\natexlab{}.
\newblock \showarticletitle{{Cognitive mechanisms in sensemaking: A qualitative
  user study}}.
\newblock \bibinfo{journal}{\emph{Journal of the Association for Information
  Science and Technology}} \bibinfo{volume}{71}, \bibinfo{number}{2}
  (\bibinfo{year}{2020}), \bibinfo{pages}{158--171}.
\newblock
\showISSN{23301643}
\urldef\tempurl%
\url{https://doi.org/10.1002/asi.24221}
\showDOI{\tempurl}


\end{thebibliography}
